\documentclass[lettersize,journal]{IEEEtran}
\usepackage{amsmath,epsfig,amssymb,verbatim}
\usepackage{cite}
\usepackage{graphicx}
\usepackage{bm,bbm}
\usepackage{xcolor}
\usepackage{hhline}
\usepackage{subfigure}
\usepackage{stackengine}
\usepackage{subcaption}
\usepackage{cleveref}
\usepackage{multicol}
\usepackage{multirow}
\def\delequal{\mathrel{\ensurestackMath{\stackon[1pt]{=}{\scriptstyle\Delta}}}}

\newcommand{\E}{\mathbb{E}}

\newcommand{\Lp}{\mathcal{L}}

\newcommand{\1}{\mathbbm{1}}

\newtheorem{Theorem}{Theorem}
\newtheorem{Lemma}{Lemma}

\begin{document}
\title{Coverage Analysis for 3D Indoor Terahertz Communication System Over Fluctuating Two-Ray Fading Channels}
\author{Zhifeng Tang,~\IEEEmembership{Member,~IEEE,}
Nan Yang,~\IEEEmembership{Senior Member,~IEEE,}\\Salman Durrani,~\IEEEmembership{Senior Member,~IEEE,} Xiangyun Zhou,~\IEEEmembership{Fellow,~IEEE,}\\
Markku Juntti~\IEEEmembership{Fellow,~IEEE,} and Josep Miquel Jornet~\IEEEmembership{Fellow,~IEEE}
\vspace{-1em}
\thanks{This work was supported by the Australian Research Council Discovery Project (DP230100878).}
\thanks{Z. Tang, N. Yang, S. Durrani, and X. Zhou are with the School of Engineering, Australian National University, Canberra, ACT 2601, Australia (Emails: \{zhifeng.tang, nan.yang, salman.durrani, xiangyun.zhou\}@anu.edu.au).}
\thanks{Markku Juntti is with the Centre for Wireless Communications, University
of Oulu, Oulu 90014, Finland (Email: markku.juntti@oulu.fi).}
\thanks{Josep Miquel Jornet is with the Department of Electrical and Computer Engineering, Northeastern University, Boston, MA 02120, USA (Email: j.jornet@northeastern.edu).}}

\maketitle

\begin{abstract}
    In this paper, we develop a novel analytical framework for a three-dimensional (3D) indoor terahertz (THz) communication system. Our proposed model incorporates more accurate modeling of wall blockages via Manhattan line processes and precise modeling of THz fading channels via a fluctuating two-ray (FTR) channel model. We also account for traditional unique features of THz, such as molecular absorption loss, user blockages, and 3D directional antenna beams. Moreover, we model locations of access points (APs) using a Poisson point process and adopt the nearest line-of-sight AP association strategy. Due to the high penetration loss caused by wall blockages, we consider that a user equipment (UE) and its associated AP and interfering APs are all in the same rectangular area, i.e., a room. Based on the proposed rectangular area model, we evaluate the impact of the UE's location on the distance to its associated AP. We then develop a tractable method to derive a new expression for the coverage probability by examining the interference from interfering APs and considering the FTR fading experienced by THz communications. Aided by simulation results, we validate our analysis and demonstrate that the UE's location has a pronounced impact on its coverage probability. Additionally, we find that the optimal AP density is determined by both the UE's location and the room size, which provides valuable insights for meeting the coverage requirements of future THz communication system deployment.
\end{abstract}


\section{Introduction}
In recent years, the increasing demand for high-speed and high-capacity data transmission has sparked significant interest in terahertz (THz) communications, positioning it as a key technology for envisioned sixth-generation (6G) wireless communications \cite{6GNet2020}. In particular, THz communications refer to the transmission and reception of signals in the THz frequency, i.e., 0.1-10 THz, which has a large amount of spectrum resources and a picosecond-level symbol duration \cite{THzComMCS2024,Elayan2020ojcom}. With its ability to offer ultrabroad bandwidth and high data rate, THz communications unlock rich opportunities for advanced applications, including ultra-high-speed data transfer in small cells and among proximal devices \cite{Shafie2023}, such as the immersive and interactive communications in the indoor environment. Despite the promise, unleashing the full benefits of THz communications poses unprecedented challenges that are not encountered at lower frequencies \cite{Ian2014}. Due to the short wavelength of the THz frequency, THz communications experience extremely high spreading loss, and the energy of THz signals is significantly attenuated and absorbed by atmospheric gas molecules, while THz waves are vulnerable to blockages \cite{Jornet2011,Han2016tsp}. These factors indicate that the environment significantly impacts the performance of THz communication systems. In order to integrate THz networks into future 6G systems, it is critical to understand the unique characteristics of THz channels, and address such characteristics in the analysis, design, and development of THz communication systems.

To evaluate the performance of communication systems, stochastic geometry has been widely used to model the location of nodes and assess the key performance metrics, such as coverage probability, which measures the coverage performance of systems \cite{haenggi2012stochastic,Andrews2011Tcom,Hmamouche2021IEEE,Guo2014tcom}. Moreover, due to the high spreading loss and molecular absorption loss in the THz frequency, directional antenna techniques are utilized to generate narrow signal beams with high directional gains, effectively compensating for the severe path loss \cite{Yu2017JSAC,Rebato2019tcom}. By leveraging stochastic geometry and incorporating directional antennas at transmitters, \cite{Kokkoniemi2017twc} analyzed the coverage probability of the THz system by considering the interference impact within a limited regime and approximating the interference as a log-logistic distribution. Additionally, \cite{Yao2019iccc} evaluated the coverage probability in indoor THz communication systems, considering both line-of-sight (LoS) and non-line-of-sight (NLoS) signal propagation. Beyond THz-only systems, integrated sub-6 GHz and THz communication systems have been explored in \cite{Sayehvand2020wcl,Lou2023,Sharma2023coml}. Specifically, \cite{Sayehvand2020wcl} derived the coverage probability of such systems, and \cite{Lou2023} introduced a hybrid sub-6 GHz and THz relay selection protocol to improve the coverage probability. Furthermore, \cite{Sharma2023coml} implemented a maximal ratio combining scheme to optimize the coverage probability by combining received signals from both sub-6 GHz and THz links.


Due to the significant penetration loss in the THz frequency, blockages can severely impact the coverage performance of THz communication systems. In indoor environments, the human body is one of the most common blockages. The presence of human bodies within the THz signal propagation path can significantly attenuate THz signals \cite{Gapeyenko2016icc}. Incorporating the effect of human blockages, \cite{Venugopal2016Access} analyzed their impact on signal propagation and evaluated the system's coverage probability. In \cite{Petrov2017}, the authors assessed the system's coverage probability by evaluating the mean and variance of the interference with a Taylor series approximation. Additionally, \cite{MOLTCHANOV201821PhysCom} provided an analytical approximation of the interference and the signal-to-interference ratio (SIR) in THz communication systems. Beyond humans, walls also cause blockages in the indoor environment. Considering the joint impact of human and wall blockages, \cite{Shafie2021JSAC} conducted a coverage analysis for a three-dimensional (3D) THz communication system. The integration of sub-6 GHz and THz communication systems was explored in \cite{Kouzayha2023twc}, where the impact of both human and wall blockages on the coverage probability and average transmission rate was evaluated. Although \cite{Shafie2021JSAC,Kouzayha2023twc} evaluated the impact of wall blockages on the system performance, they modeled walls using a Boolean scheme of straight lines, which may not accurately reflect actual wall deployment. To more precisely represent wall deployment, \cite{Wu2021TWC} employed the Manhattan Poisson line process (MPLP) as the wall blockage model, dividing the indoor environment into multiple rectangular spaces. However, due to the complexity of analyzing system performance within these rectangular spaces, the authors approximated them as circular areas to simplify the coverage analysis of THz communication systems.


Although THz channels heavily rely on the LoS component and suffer from significant reflection, diffraction, and scattering losses, small-scale fading still exists in THz channels due to atmospheric aerosols acting as scatterers \cite{Boulogeorgos2019pimrc}. To characterize the complex signal propagation environment with greater flexibility, \cite{Ye2022tcom} employed the Nakagami-$m$ distribution to model the small-scale fading gain in THz channels and analyzed its impact on system's coverage. Recent studies have further evaluated the small-scale fading in THz channels \cite{Papasotiriou2021,Romero2017twc}. Specifically, \cite{Papasotiriou2021} conducted three measurements at the center frequency of 143.1 GHz with a total bandwidth of 4 GHz. The measurement results indicated that the $\alpha$-$\mu$ distribution provides a more accurate fit for characterizing the small-scale fading of THz channels, compared to traditional Rayleigh, Rice, and Nakagami-$m$ distributions. Similarly, \cite{Romero2017twc} conducted THz channel measurements at a train test center, showing that the fluctuating two-ray (FTR) distribution offers a more accurate fit than Gaussian, Rician, and Nakagami-$m$ distributions. To further refine the modeling of small-scale fading in THz channels, \cite{Du2022tcom} compared the $\alpha$-$\mu$ and FTR distributions, concluding that the FTR distribution more accurately characterizes the small-scale fading in THz channels. By adopting the FTR distribution to model the small-scale fading gain of THz channels, \cite{Sharma2024twc} explored the secrecy reliability of THz systems, highlighting the importance of precise small-scale fading models for accurate performance analysis.


\textit{Paper Contributions:} In this paper, we investigate the coverage probability of downlink transmission in an indoor THz communication system. We evaluate the impact of different types of blockages, directional antennas, and interference on the coverage probability of THz communication systems. The main contributions of this work are summarized as follows.
\begin{itemize}
    \item We characterize the joint impact of wall and human blockages on THz communication systems. Specifically, we model wall blockages using Manhattan line processes (MLPs) for more accurate representation compared to the existing methods, while modeling human blockages using random cylinder processes, consistent with the existing works. 
    Due to the high penetration loss caused by wall blockages, a user equipment (UE) and its associated associate point (AP) and interfering APs are assumed to be in the same room. We then analyze how the UE's location in a room affects its distance to its associated AP and the intensity of interfering APs. Our analytical results show that the use of the existing circular area model for analyzing the association distance between a UE and its AP is only accurate when the UE is located closer to the center of the room but the accuracy deteriorates for a UE located closer to the corner of the room. Additionally, a UE located in the corner experiences less interference from interfering APs than a UE at the center. These findings reveal how the UE's location and the room size affect the system performance, and these impacts have not been explored in previous studies.

    \item We consider the effect of 3D directional antennas and derive the hitting probability of an interfering AP being within the antenna beam of a UE. Our analytical results demonstrate that the hitting probability monotonically increases as the distance from the UE to its associated AP increases and varies depending on the UE's location in the room. These variations are not captured by prior studies that employed a circular area model. We then develop a tractable analytical method, using tools from stochastic geometry, to characterize the coverage probability of the THz communication system. Specifically, we derive the expression for the coverage probability by analyzing the intensity of interfering APs in the room and considering that the THz communication link experiences FTR fading. Our analysis is validated through simulation results.
    
    \item Our results theoretically show how the coverage probability is affected by the UE's location and the room size. Specifically, a UE at the center experiences a higher coverage probability in a small room, but a lower coverage probability in a large room, compared to a UE in the corner. Moreover, the coverage probability derived for the circular area model is close to that of a UE at the center of the room but provides a poor approximation for a UE in the corner. It implies that the analysis for a circular area model can be adapted to simplify the analysis for UEs at the center, but not suitable for the UE in the corner. Furthermore, our investigation into the optimal AP density for coverage probability maximization shows that the UE's location has a significant impact on the optimal AP density, indicating that both room size and the UE's location are determinants. These insights are essential for designing future THz systems in indoor environments to meet coverage requirements.

\end{itemize}

The rest of the paper is organized as follows. In Section \ref{Sec:System}, the system development, directional antenna model, and THz channel model are described. In Section \ref{Sec:Coverage}, we first evaluate the AP association, the interfering AP intensity, and the impact of directional antennas, and then derive the coverage probability. Numerical and simulation results are provided in Section \ref{Sec:Num}, and the paper is concluded in Section \ref{Sec:Conclusion}.

\section{System Model}\label{Sec:System}
\begin{figure}[t]
    \centering
    \includegraphics[width=0.95\columnwidth]{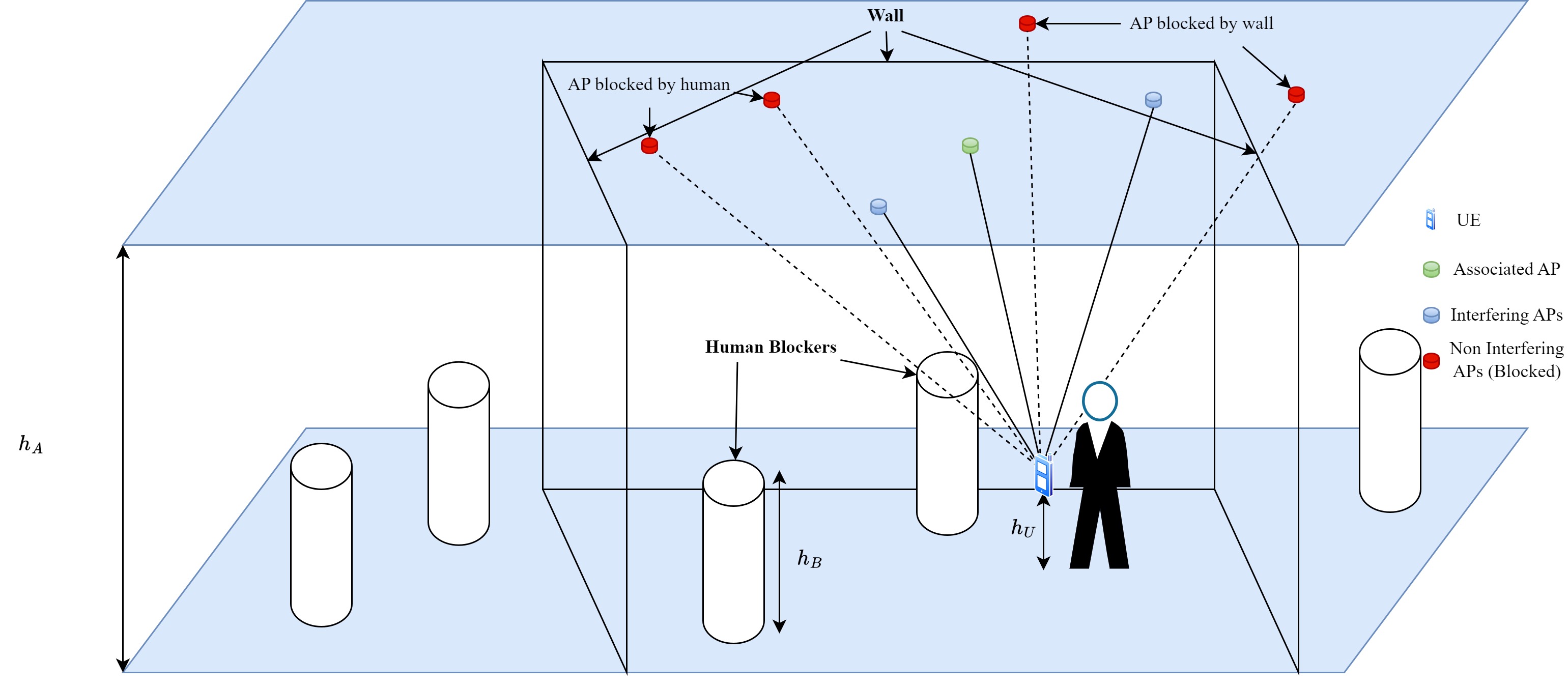}
    \caption{Illustration of the 3D indoor THz communication system where a typical UE associates with a non-blocked (green) AP in the presence of interfering (blue) APs. The non-interfering APs (red) include those blocked by human and wall blockers.}
    \vspace{-1.5em}
    \label{fig:system_model}
\end{figure}

This work considers a generalized 3D indoor THz communication system, illustrated in Fig.~\ref{fig:system_model}. In this system, multiple APs mounted on the ceiling transmit THz signals to UEs. The following subsections provide a detailed description of the considered system.

\subsection{System Deployment}

We assume that THz APs are mounted on the ceiling with a fixed height $h_A$ in the indoor environment and their locations follow a Poisson point process (PPP) with the density of $\lambda_A$. We also assume that UEs, having a fixed height $h_U$, are randomly distributed on the ground. Within this system, we randomly select one UE and refer to it as the typical UE, denoted by $U_0$.

\begin{figure*}[ht]
    \centering
        \subfigure[The top view of a floor in a building.]{\label{fig:Topview}\includegraphics[width=0.3\textwidth]{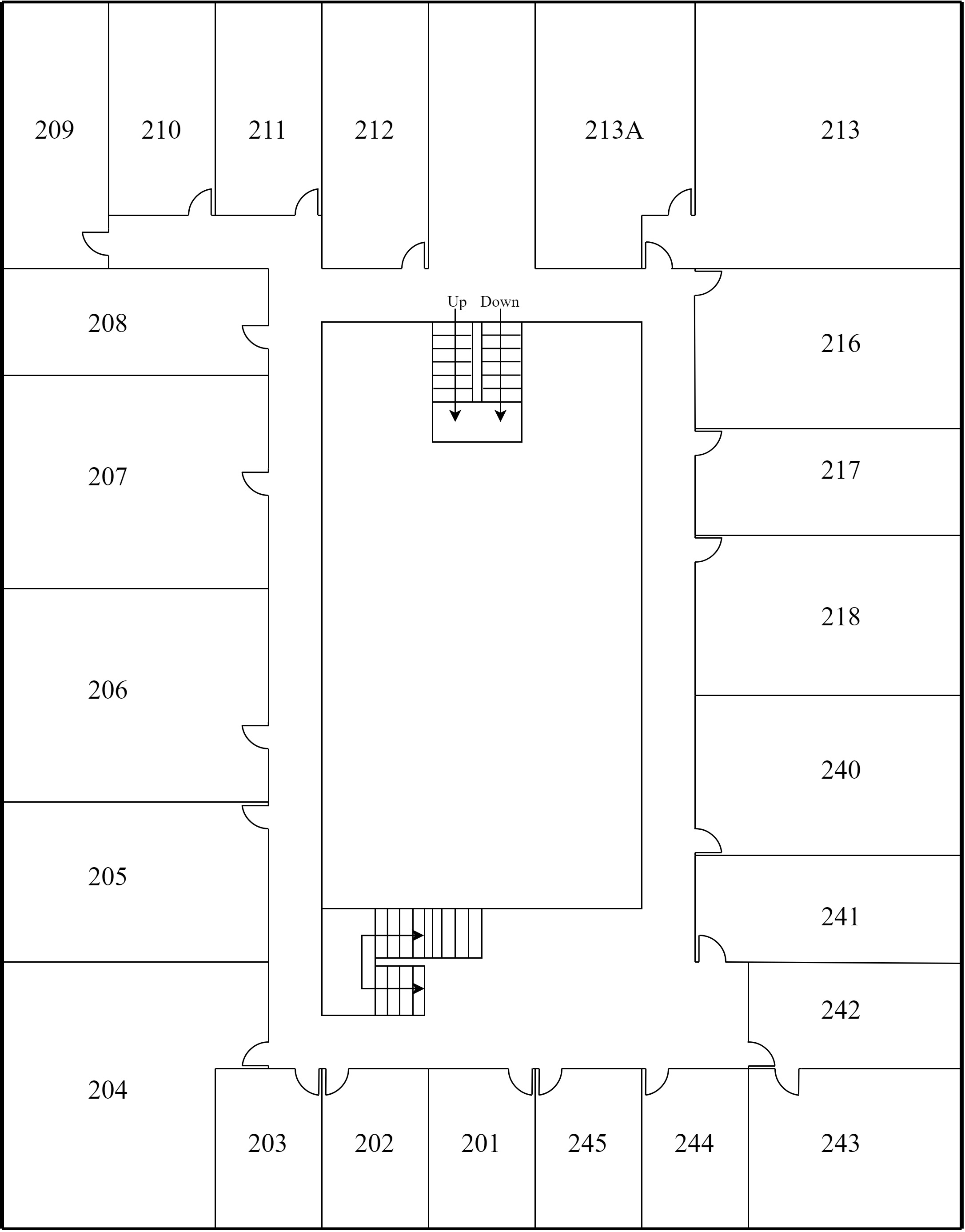}}
        \subfigure[MLPs wall model.]{    \label{fig:MLP}        
            \includegraphics[width=0.3\textwidth]{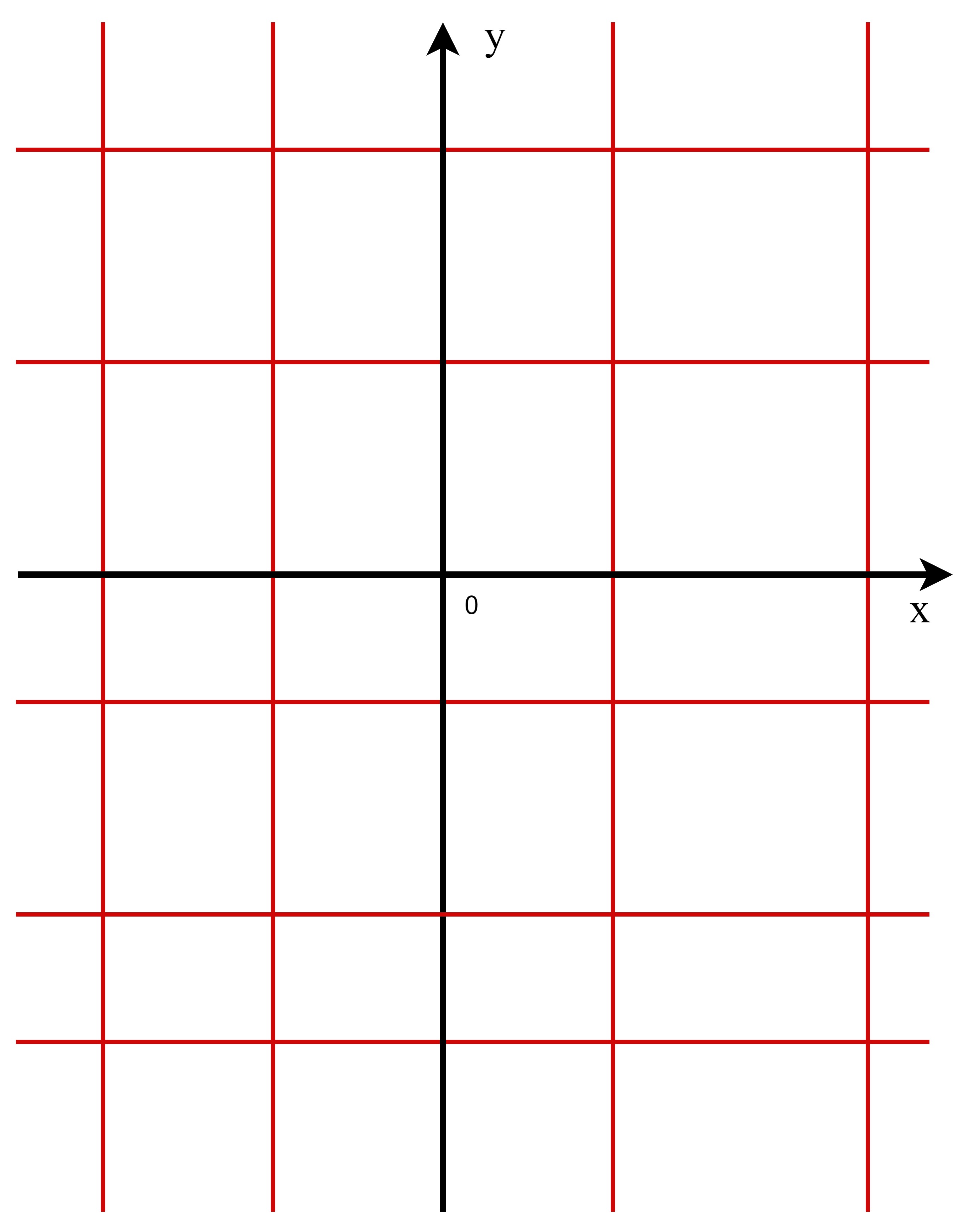}}
        \subfigure[The room $\mathbf{R}$ where $U_0$ is located.]{  \label{fig:Room1}        \includegraphics[width=0.3\textwidth]{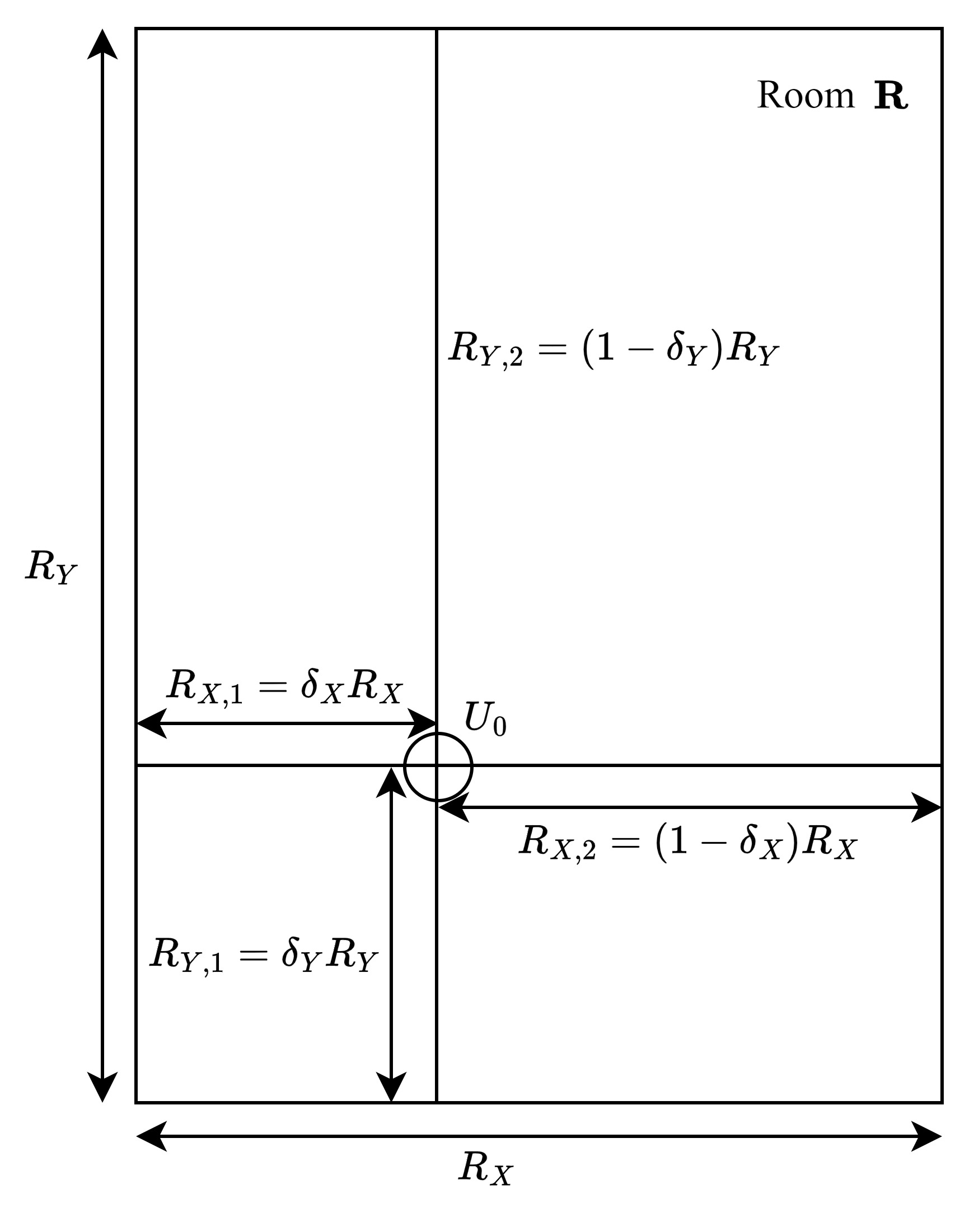}}
        \caption{The wall deployment model of the THz communication system.}
        \vspace{-1.5em}
        \label{fig:WallDeployment}
\end{figure*}

Consider the top view of a typically indoor building environment shown in Fig.~\ref{fig:Topview}. We can see that architecturally it is typically divided into several rectangular subspaces, e.g., rooms, by pairs of perpendicular walls. In practice, these walls may be flexible or temporary walls which allow the room geometry to be easily altered by adding or removing the walls. Inspired by this, we employ MLPs to describe the wall blockage model. Specifically, walls are oriented at either $0$ or $\pi/2$ angles to ensure that they are parallel or perpendicular to each other. The walls are modeled as two independent MLPs, whose centers are distributed along x-axis and y-axis, as illustrated in  Fig.~\ref{fig:MLP}. Based on this wall model, we denote the room where $U_0$ is located as $\mathbf{R}$, with $\mathbf{R}=\{R_X,R_Y\}$ representing the width and length of the room, respectively, as illustrated in Fig.~\ref{fig:Room1}. We also denote $\mathbf{R}_{U_0}=\{R_{X,1},R_{X,2},R_{Y,1},R_{Y,2}\}$ as the horizontal distance from the typical UE, $U_0$, to its left, right, front, and rear walls, respectively. Since $R_X=R_{X,1}+R_{X,2}$ and $R_Y=R_{Y,1}+R_{Y,2}$, the location of $U_0$ in the room $\mathbf{R}$ can be expressed by $\mathbf{\delta}=\{\delta_X,\delta_Y\}$, where $\delta_X=\frac{R_{X,1}}{R_X}$ and $\delta_Y=\frac{R_{Y,1}}{R_Y}$. In the indoor environment, THz signal transmission through walls is neglected due to the very high penetration loss caused by wall blockages \cite{Shafie2021JSAC}. Consequently, the typical UE, $U_0$, its associated AP, and interfering APs are all within the same room, $\mathbf{R}$. 

\begin{figure}[ht]
    \centering
    \captionsetup[subfloat]{labelfont=footnotesize,textfont=footnotesize}
        \subfigure[The top view.]{
            \includegraphics[width=0.45\columnwidth]{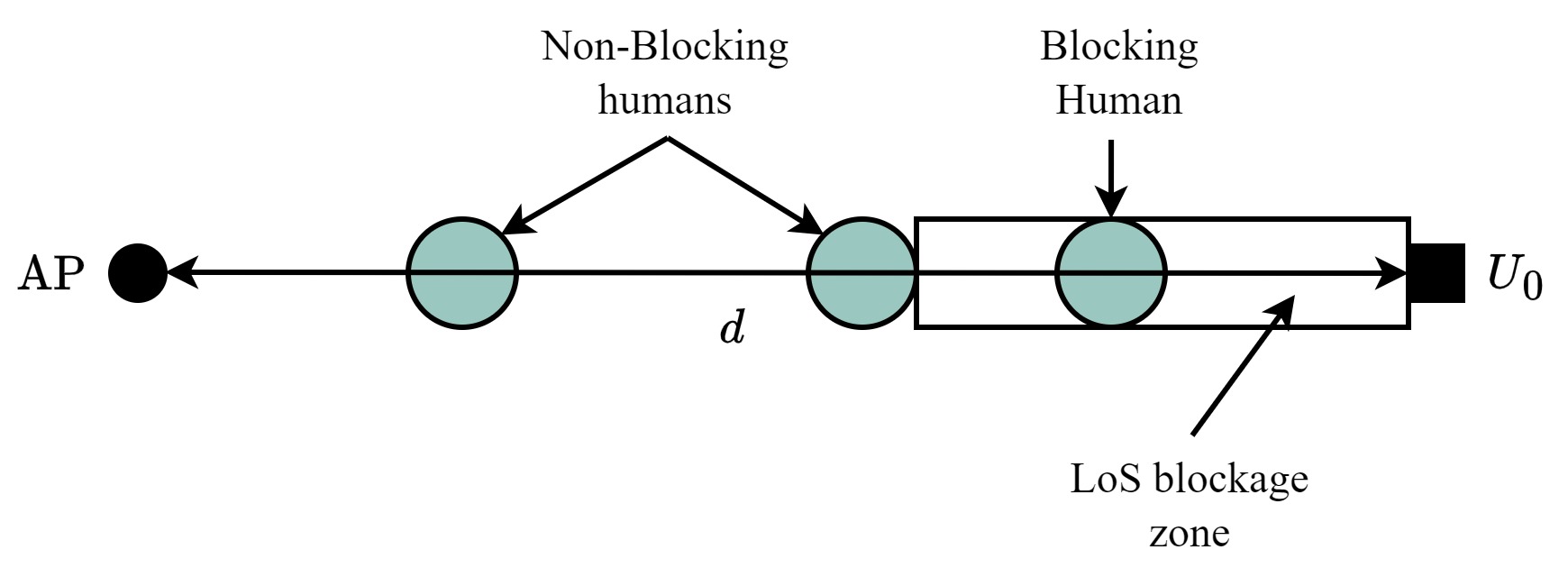}
            }\label{fig:TopviewHB}
        \subfigure[The vertical view.]{
            \includegraphics[width=0.45\columnwidth]{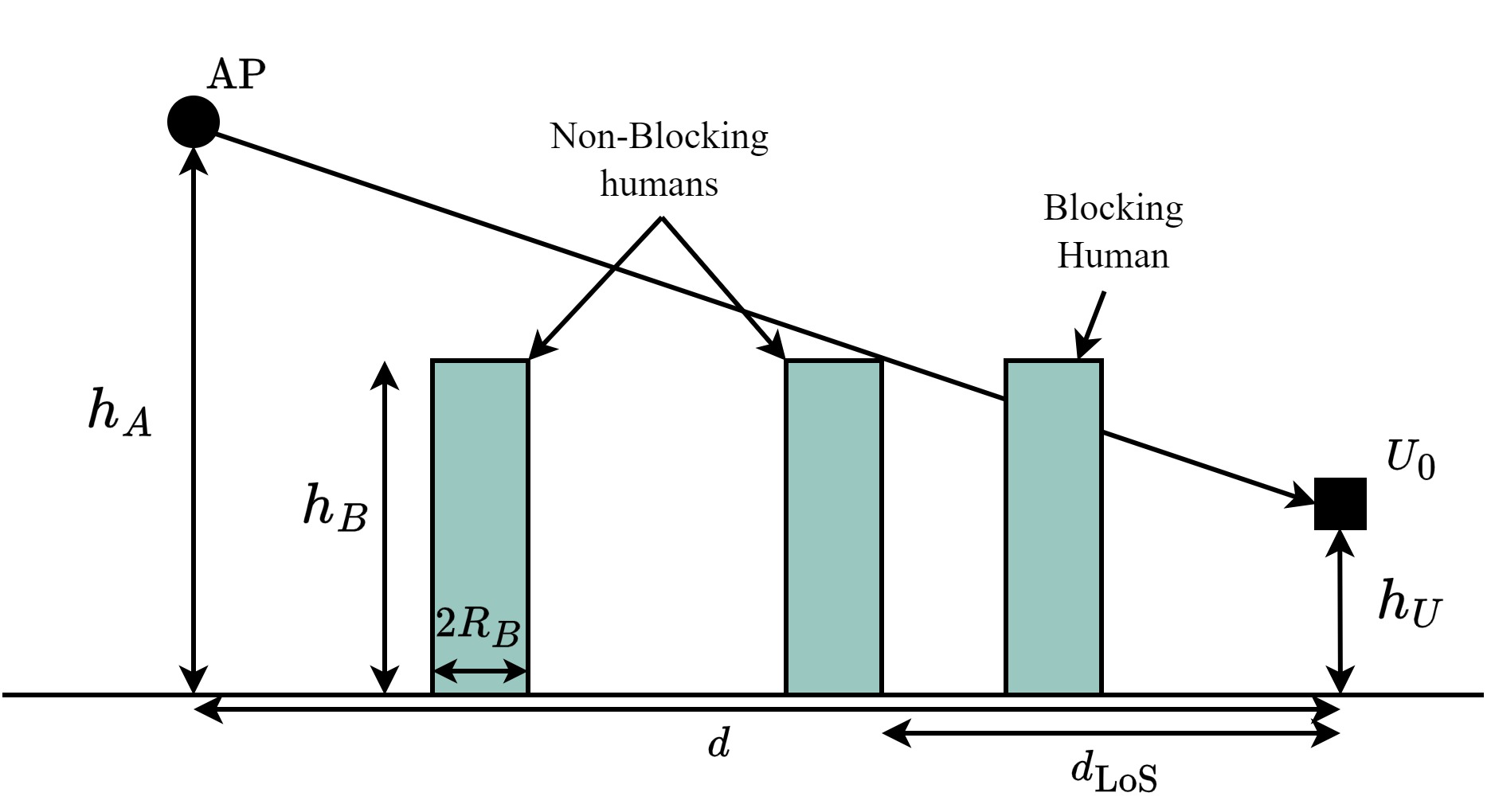}
            }\label{fig:VertviewHB}
        \caption{Top and vertical views of human blockage for an AP-UE link.}
        \label{fig:HB}
\end{figure}

Apart from the wall blockage, another blockage is caused by human bodies. Following the state-of-the-art studies, e.g., \cite{Kouzayha2023twc}, we model each human body as a cylinder with a radius of $R_B$ and a height of $h_B$. We model the bottom center of cylinders by a 2D-PPP with the density $\lambda_B$. As shown in Fig.~\ref{fig:HB}, if the bottom center of a human body is within the LoS blockage zone, the AP-UE signal transmission link is blocked. Conversely, if a human body is entirely outside the LoS blockage zone, the AP-UE signal transmission link is considered as LoS, and the signal transmission from the AP can be received by the UE. According to \cite{Wu2021TWC}, the probability that an AP-UE link is blocked by a human body is given by $p_B(d) = \exp(-\alpha d)$, where $\alpha \delequal 2\lambda_B R_B (h_B-h_U)/(h_A-h_U)$ and $d$ is the horizontal distance between the AP and the UE.

In the considered network, we adopt the nearest-LoS-AP association strategy \cite{Wu2021TWC}, i.e., each UE associates with its nearest AP that has a LoS propagation path to the UE. For the typical UE, we denote its associated AP by $AP_0$ and the horizontal distance between $U_0$ and $AP_0$ by $d_0$. For other LoS non-associated AP, we denote $AP_{i}$ as the $i$th nearest LoS non-associated AP of $U_0$ and $d_i$ as the distance between $U_0$ and $AP_i$, where $\Psi_{AP} = \{AP_1,AP_2,\cdots\}$.



\subsection{Antenna Model}


In this work, we assume the use of 3D directional antennas to enhance signal strength to compensate for the severe path loss in THz propagation \cite{Yuan2020}. The directional 3D beams are approximated by a 3D pyramidal-plus-sphere sectored antenna model, as shown in \cite[Fig.~2]{Shafie2021JSAC}. 
Specifically, APs and UEs have different antenna parameters due to the different hardware requirements. Given that APs are usually equipped with a larger number of antennas, they are able to produce a narrower beamwidth and higher antenna gain compared to UEs. Based on the principles of the antenna theory \cite{Shafie2021JSAC}, the antenna gains of the main lobe and the side lobes for APs and UEs are expressed as
\begin{align}
    G_{\Xi}^m = \frac{\pi }{(1+k_{\Xi})\arcsin(\tan(\phi_{\Xi,H}/2)\tan(\phi_{\Xi,V}/2))}
\end{align}
and
\begin{align}
    G_{\Xi}^s = \frac{\pi k_{\Xi}}{(1+k_{\Xi})(\pi-\arcsin(\tan(\phi_{\Xi,H}/2)\tan(\phi_{\Xi,V}/2)))},
\end{align}
respectively, where $\Xi\in\{\mathrm{A},\mathrm{U}\}$ with $\mathrm{A}$ representing AP and $\mathrm{U}$ representing UE, $k_{\Xi}$ is the ratio of the power fraction concentrated along the side lobes to the fraction of power concentrated along the main lobe, and $\phi_{\Xi,H}$ and $\phi_{\Xi,V}$ are the horizontal and vertical widths of beam, respectively.

Since the antenna beams of each UE and its associated AP are directed towards each other, the transmit and receive antenna gains of the received signal at $U_0$ are equal to their main lobe gains, given by
\begin{align}
    G_{0,A} = G_{A}^m \text{ and } G_{0,U} = G_{U}^m.
\end{align}
For other LoS non-associated APs, we denote $G_{i,A}$ and $G_{i,U}$ as the transmit antenna gain of $AP_i$ and the receive antenna gain at $U_{0}$ from $AP_i$, respectively.

\subsection{THz Channel Model}

The signal propagation at THz frequencies is affected by the distance-dependent large-scale fading and the multipath-induced small-scale fading. In THz communication systems, the large-scale fading is primarily determined by the spreading loss and the molecular absorption loss. The average received power at $U_0$ from $AP_i$ located at the distance $d_i$ in the 3D indoor environment is expressed as
\begin{align}
    \overline{P_i} = g_{i}W(d_{i}),
\end{align}
where $g_{i}\delequal \frac{P_t G_{i,A}G_{i,U} c^2}{(4\pi f)^2}$, $P_t$ is the transmit power, $G_{i,A}$ and $G_{i,U}$ are the effective antenna gains at $AP_i$ and $U_0$, respectively, for the $AP_{i}$-$U_{0}$ link, $c=3\times10^8$ m/s is the light speed, $f$ is the operating frequency, $W(d_{i})=\frac{1}{d_{i}^2+(h_A-h_U)^2}\exp\left(-\epsilon(f)\sqrt{d_{i}^2+(h_A-h_U)^2}\right)$, and $\epsilon(f)$ is the molecular absorption coefficient of frequency $f$. Here, $\frac{c^2}{(4\pi f)^2(d_{i}^2+(h_A-h_U)^2)}$ represents the spreading loss and $\exp\left(-\epsilon(f)\sqrt{d_{i}^2+(h_A-h_U)^2}\right)$ represents the molecular absorption loss.

Compared to Rayleigh, Rician, Nakagami-$m$, and $\alpha-\mu$ distributions, recent measurements have shown that the FTR distribution is particularly well-suited for THz communications, due to the dominance of LoS and single reflected paths accompanied by diffraction and scattering at THz frequencies for short-distance transmission \cite{Papasotiriou2021}. Hence, in this work, we denote $H_i$ as the small-scale fading gain for the $AP_i$-$U_0$ link and model $H_i$ using the FTR distribution. Under this modeling, the probability density function (PDF) and cumulative distribution function (CDF) of $H_i$ are given by
\begin{align}\label{eq:pdfFTR}
    f_{H_i}(h) =  \frac{m^m}{\Gamma(m)}\sum\limits_{j=0}^{\infty}\frac{K^j r_j}{j!\Gamma(j+1)(2\sigma^2)^{j+1}}h^je^{-\frac{h}{2\sigma^2}}
\end{align}
and
\begin{align}\label{eq:cdfFTR}
    F_{H_i}(h) 
    &=1-\frac{m^m}{\Gamma(m)}\sum\limits_{j=0}^{\infty}\frac{K^j r_j}{j!\Gamma(j+1)}\Gamma\left(j+1,\frac{h}{2\sigma^2}\right),
\end{align}
respectively \cite{Du2022tcom}. In \eqref{eq:pdfFTR} and \eqref{eq:cdfFTR}, $\Gamma(\cdot)$ is the gamma function, while $\gamma(\cdot,\cdot)$ and $\Gamma(\cdot,\cdot)$ are the lower and upper incomplete gamma functions, respectively. The parameter $m$ denotes the severity of fading, $K$ represents the average power ratio of the dominant component to the remaining diffuse multipath, $2\sigma^2$ is the average power of the diffuse component over FTR fading, and $r_j$ is given by
\begin{align}
    r_j &= \sum\limits_{k=0}^j {j\choose k} \sum\limits_{l=0}^k{k\choose l}\Gamma(j\!+\!m\!+\!2l\!-\!k)\left(m\!+\!K\right)^{-(j\!+\!m\!+\!2l\!-\!k)}\notag\\
   & \times K^{2l-k}\left(\frac{\Delta}{2}\right)^{2l}(-1)^{2l-k}\Omega_{j+m-l}^{k-2l}\left(\left[\frac{K\Delta}{m+k}\right]^2\right),
\end{align}
where $\Delta$ is the parameter representing dominant waves similarity and $\Omega^{\mu}_{\upsilon}(x)$ is defined as
\begin{align}\label{eq:RMUV}
&\Omega^{\mu}_{\upsilon}(x) =\notag\\
&\left\{
\begin{aligned}
    &\left(\frac{\upsilon-\mu}{2}\right)_{\mu}\left(\frac{\upsilon-\mu+1}{2}\right)_{\mu} \frac{x^{\mu}}{\mu !}\\
    &\times _2F_1\left(\frac{\upsilon\!+\!\mu}{2},\frac{\upsilon\!+\!\mu\!+\!1}{2};1\!+\!\mu;x\right), &\text{if }\mu \in \mathbb{N}^+,\\
    &\frac{_2F_1\left(\frac{\upsilon\!-\!\mu}{2},\frac{\upsilon\!-\!\mu\!+\!1}{2};1-\mu;x\right)}{\Gamma(1-\mu)},   &\text{otherwise}.
\end{aligned}
\right.
\end{align}
In \eqref{eq:RMUV}, $_2F_1(\cdot,\cdot;\cdot;\cdot)$ is the Gauss hypergeometric function and $(\cdot)_a$ is the Pochhammer symbol. Combining large-scale and small-scale fading models, the received power at $U_0$ from $AP_i$ is expressed as
\begin{align}
    P_i = \overline{P_i}H_{i} =  g_{i}W(d_{i})H_{i}.
\end{align}

To analyze the downlink coverage performance of a THz communication system, we employ the coverage probability, denoted by $P_c$, as our performance metric. We define it as the probability that the received signal-to-interference-plus-noise ratio (SINR) at $U_0$ exceeds a given threshold $\beta$, i.e., $P_c = \mathrm{Pr}(\mathrm{SINR}>\beta)$. Accordingly, the SINR of $U_0$ is formulated as
\begin{align}\label{eq:SINRequation}
    \mathrm{SINR} &= \frac{P_{0}}{I + N_0}=\frac{P_{0}}{\sum\limits_{AP_i\in \Psi_{AP}} P_{i} + N_0} \notag\\
    &= \frac{g_{0}W(d_{0})H_{0}}{\sum\limits_{AP_i\in \Psi_{AP}} g_{i}W(d_{i})H_{i} + N_0},
\end{align}
where $I$ represents the interference power from all LoS non-associated APs and $N_0$ denotes the noise power.

\section{Coverage Analysis}\label{Sec:Coverage}

In this section, we analyze the coverage probability of $U_0$. We note that the coverage probability can be calculated by
\begin{align}\label{eq:Pc}
    P_c \!= \!\mathrm{Pr}(\mathrm{SINR}\!>\!\beta)\! =\! \int_{0}^{\infty}\mathrm{Pr}(\mathrm{SINR}\!>\!\beta|d_0)f_{D_0}(d_0)\mathrm{d}d_0,
\end{align}
where $f_{D_0}(d_0)$ is the PDF of the horizontal distance from $U_0$ to $AP_0$. To obtain the coverage probability of $U_0$, we first derive $f_{D_0}(d_0)$ and evaluate the interfering AP intensity, which serves as the foundation for the interference analysis.

\subsection{AP Association and Interfering AP Intensity}
\begin{figure}[ht]
    \centering
        \subfigure[Angle and length of the arc in the horizontal plane in the room $\mathbf{R}$.]{\label{fig:RoomR}
             \includegraphics[width=0.45\columnwidth]{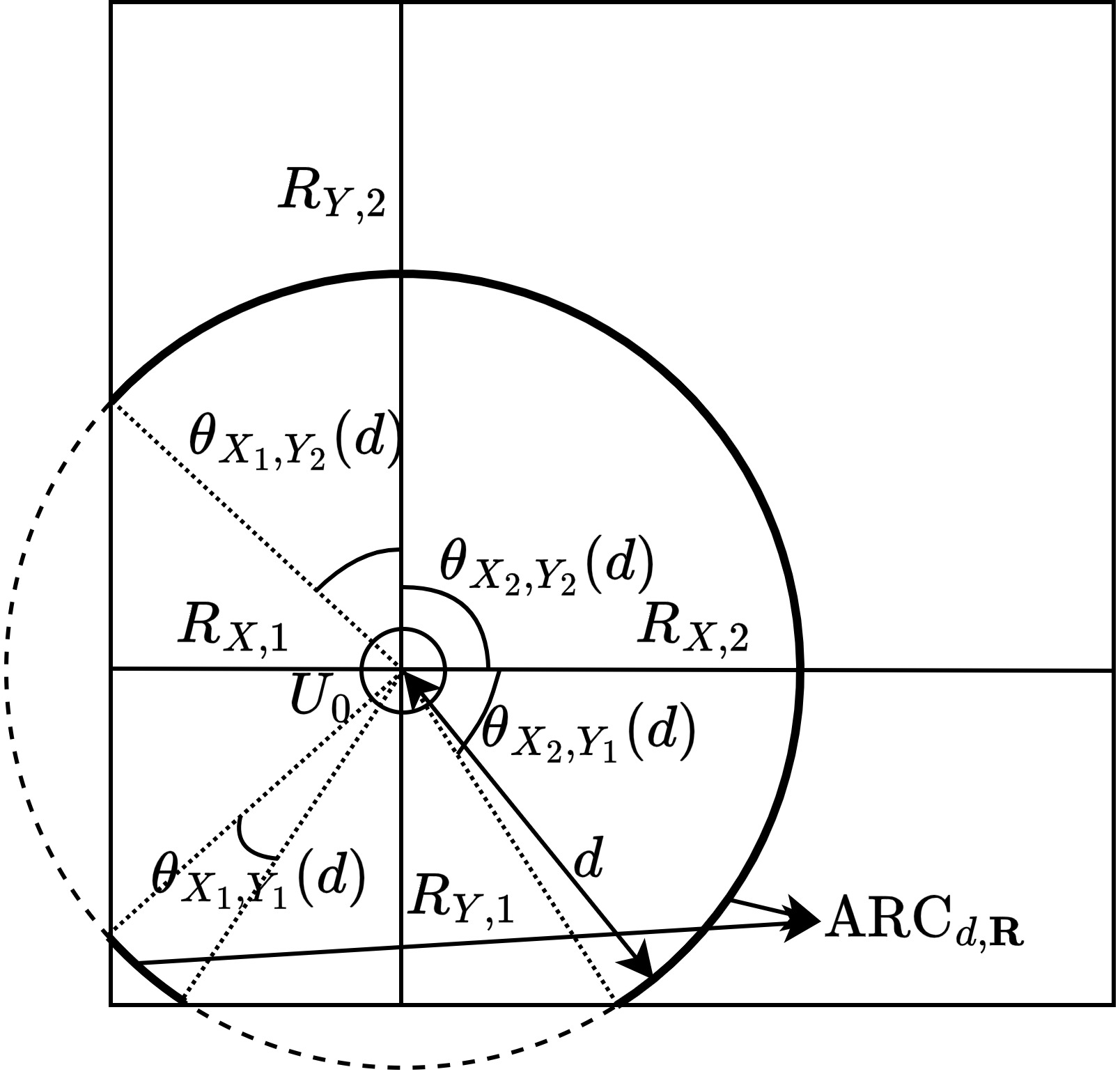}
            }
        \subfigure[Top view of one rectangular space with the length $R_{X,1}$ and width $R_{Y,1}$.]
        {    \label{fig:RoomOnePart}        
            \includegraphics[width=0.45\columnwidth]{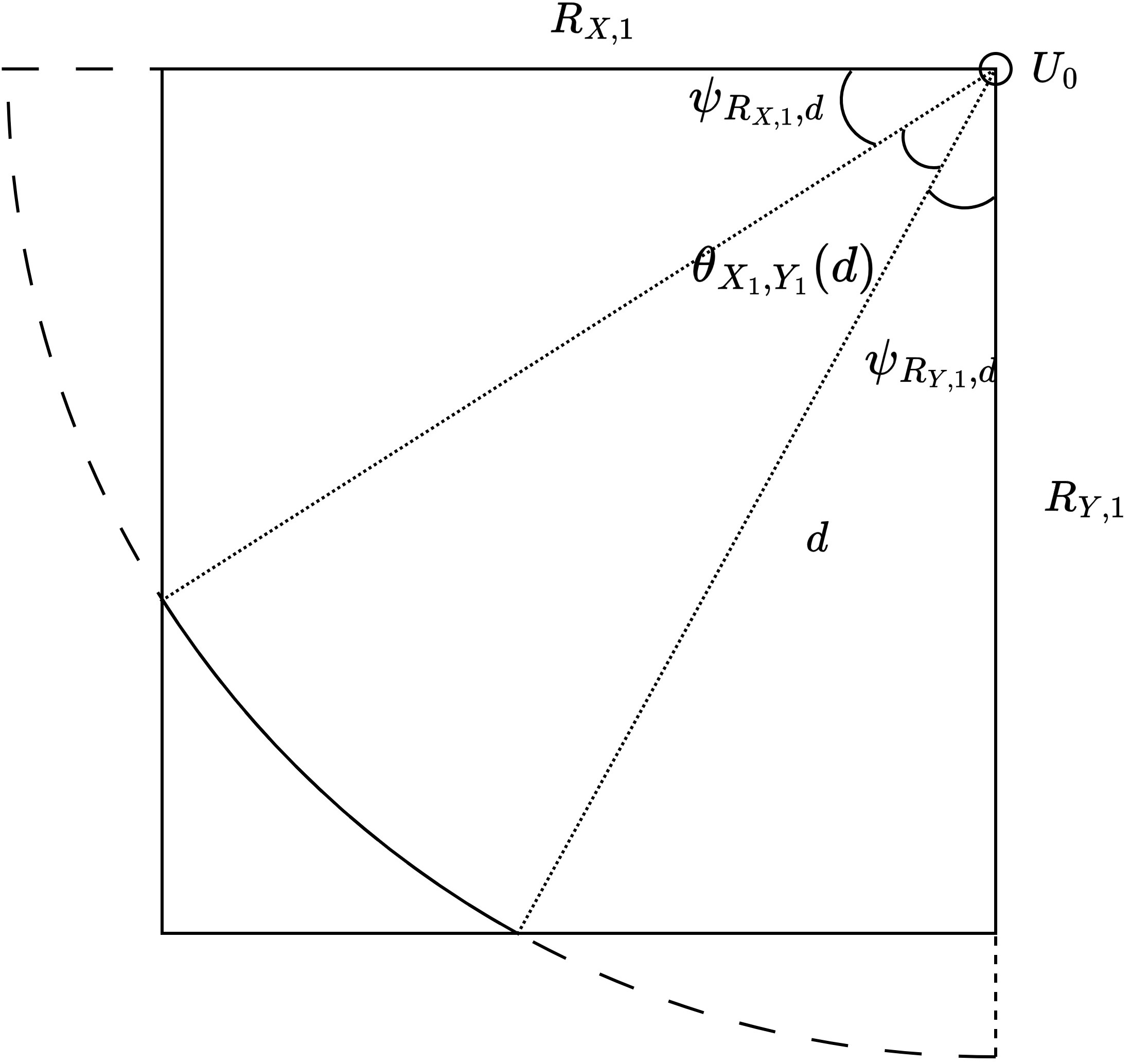}
            }
        \caption{An example of the arc with the center $U_0$ and radius $d$ in the horizontal plane in the room $\mathbf{R}$.}
        \vspace{-1.5em}
        \label{fig:Roomx}
\end{figure}

In this subsection, we derive $f_{D_0}(d_0)$ and evaluate the interfering AP intensity in the room $\mathbf{R}$. To this end, we first analyze the angle and length of the arc with the center $U_0$ and radius $d$ in the horizontal plane in the room $\mathbf{R}$, as shown in Fig.~\ref{fig:RoomR}. Here, we denote the arc with the center $U_0$ and radius $d$ in the horizontal plane in the room $\mathbf{R}$ by $\mathrm{ARC}_{d,\mathbf{R}}$.

\begin{Lemma}\label{Lemma:arc}
    The angle and length of the arc with the center $U_0$ and radius $d$ in the horizontal plane in the room $\mathbf{R}$, $\mathrm{ARC}_{d,\mathbf{R}}$, are derived as
    \begin{align}\label{eq:arcradius}
         \theta(d) = \sum\limits_{i\in\{1,2\}}\sum\limits_{j\in\{1,2\}} \left(\frac{\pi}{2}-\psi_{R_{X,i},d}-\psi_{R_{Y,j},d}\right)^{+}
    \end{align}
    and
    \begin{align}\label{eq:arclength}
        L(d) = \theta(d)d,
    \end{align}
    respectively, where $(z)^+ = \max{(z,0)}$ and
    \begin{align}\label{eq:psiab}
    \psi_{a,b} = \left\{
    \begin{aligned}
        &\arccos\left(\frac{a}{b}\right),&\text{if } b>a,\\
        &0, &\text{otherwise.}
    \end{aligned}\right.
    \end{align}
\begin{IEEEproof}   
   See Appendix~\ref{Appendix:Lemma1}.    
\end{IEEEproof}

\end{Lemma}



Based on Lemma~\ref{Lemma:arc}, the length of $\mathrm{ARC}_{d,\mathbf{R}}$ is determined by the location of $U_0$. This affects the AP intensity at the horizontal distance $d$ from $U_0$, and subsequently affects the horizontal distance between $U_0$ and $AP_0$. According to the property of the PPP, the AP intensity at a horizontal distance $d$ from $U_0$ is given by
\begin{align}
    \Lambda_{A}(d) = \lambda_{A} L(d).
\end{align} 
We then analyze the PDF of $d_0$ in the following Lemma. 
\begin{Lemma}\label{Lemma:intensity}
    The PDF of the horizontal distance between $U_0$ and $AP_0$, $d_0$, is derived as
    \begin{align}\label{eq:pdfLoSAP}
        f_{D_0}(d_0) = \Lambda_{A,\mathrm{LoS}}(d_0) e^{-\int_{0}^{d_0}\Lambda_{A,\mathrm{LoS}}(d) \mathrm{d}d},
    \end{align}
    where $\Lambda_{A,\mathrm{LoS}}(d_0) = \Lambda_{A}(d_0)e^{-\alpha d_0 }$.
    \begin{IEEEproof}
    See Appendix~\ref{Appendix:Lemma2}.
    \end{IEEEproof}
\end{Lemma}

\textit{Remark 1:} We note that the previous studies on analyzing the coverage probability of THz communication systems ignored the impact of the location of a UE on the distance to its associated AP \cite{Lou2023,Wu2021TWC}. However, by examining \eqref{eq:pdfLoSAP} and \eqref{eq:FD0}, we find that the horizontal distance between $U_0$ and $AP_0$ highly depends on the location of $U_0$ in the room, particularly for a small AP density. As $U_0$ moves from the center to the corner of the room, the CDF of $d_0$, $F_{D_0}(d_0)$, is monotonically non-increasing. This indicates that the average horizontal distance from a UE in the corner of the room to its associated AP is greater than that for a UE at the center. Consequently, the coverage performance varies for UEs at different locations. This will be further elaborated in the results in Section~\ref{Sec:Num}.

\subsection{Impact of Directional Antennas}

In this subsection, we analyze antenna gains for interfering signals, laying the foundations for the interference analysis. The antenna gain of the transmit signal from $AP_i$ to $U_0$ is the product of transmit and receive antenna gains, expressed as $G_i=G_{i,A}G_{i,U}$. Thus, we will examine transmit and receive antenna gains separately.

\subsubsection{Transmit Antenna Gain}
\begin{figure}[t]
    \centering
    \subfigure[The vertical view.]
        {    \label{fig:APbeamV}        
            \includegraphics[width=0.4\columnwidth]{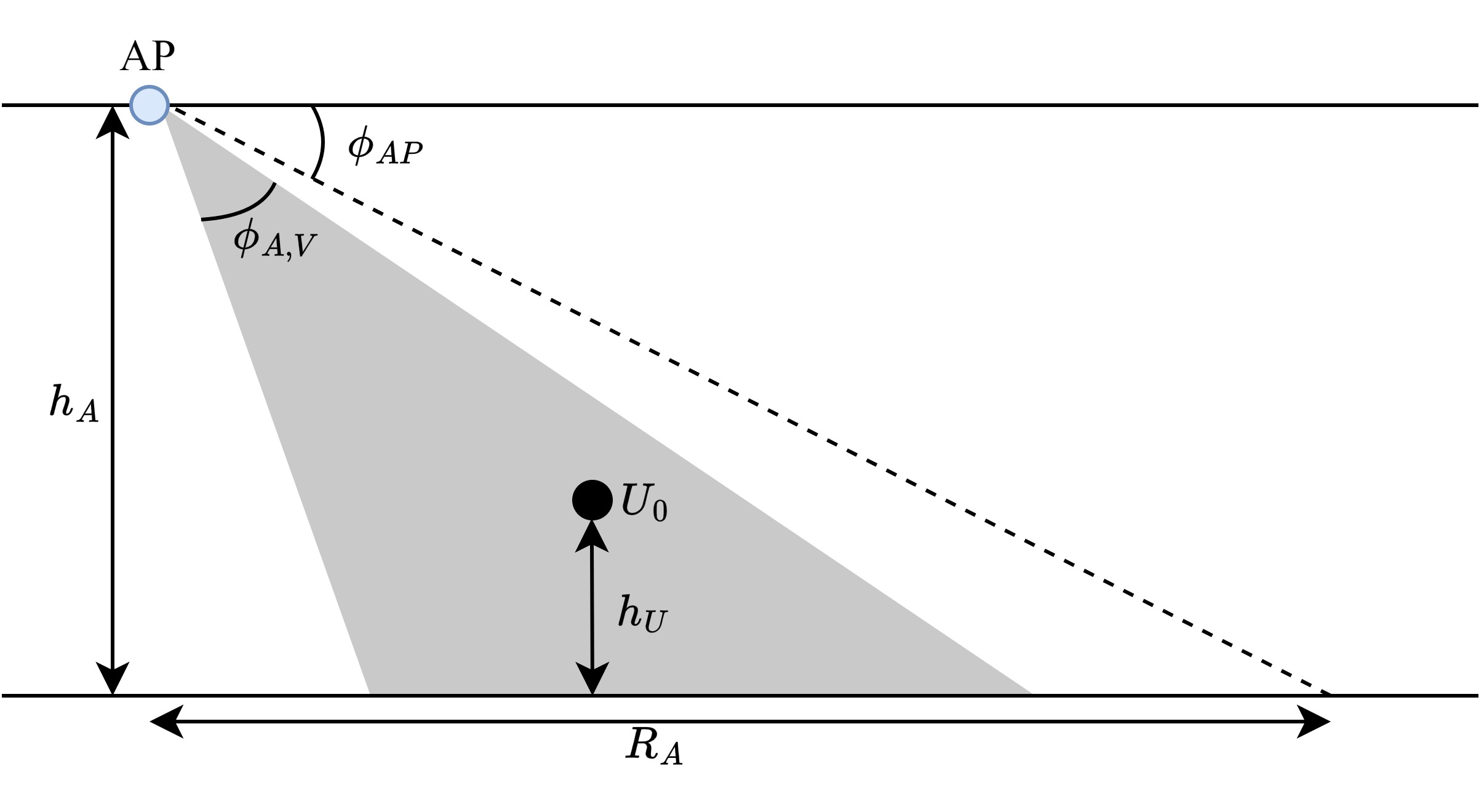}
            }
        \subfigure[The top view.]{\label{fig:APbeamH}
             \includegraphics[width=0.4\columnwidth]{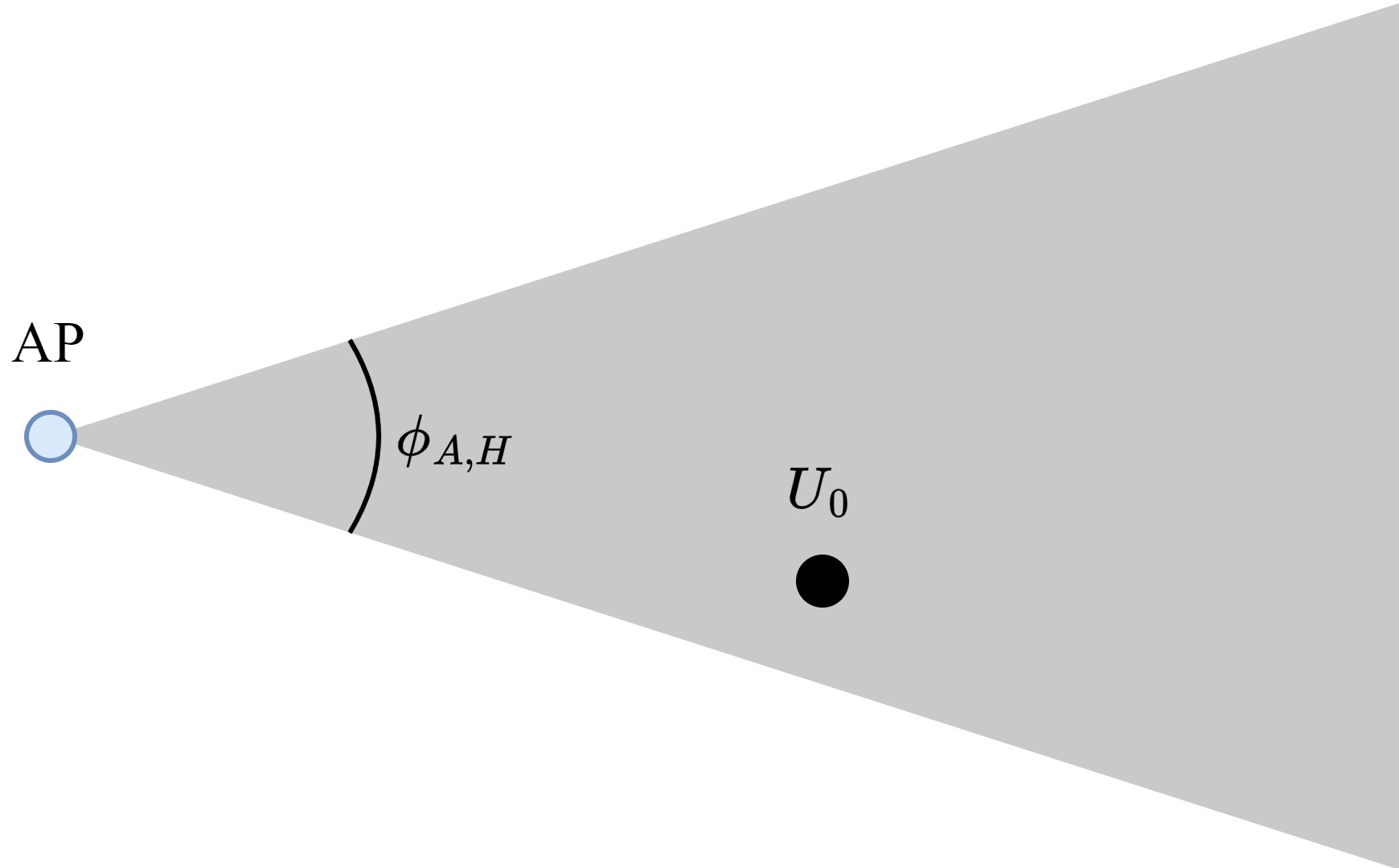}
            }
        \caption{The antenna beam of an interfering AP.}
        \vspace{-1.5em}
        \label{fig:APantenna}
\end{figure}

The transmit antenna gain depends on whether $U_0$ is within the antenna beam of the interfering AP. We assume that the depression angle from the AP is uniformly distributed over $[\phi_{AP},\pi/2]$, where $\phi_{AP} = \arctan\left(\frac{h_A-h_U}{R_A}\right)$ is the depression angle from the AP to its coverage boundary with radius $R_A$. We also assume that the horizontal beam direction from the AP is uniformly distributed over $[0,2\pi)$\footnote{We clarify that the antenna beams of APs should be directly towards their associated UEs, where the antenna beam direction of each AP depends on its location and the location of its associated UE or UEs. However, it is extremely challenging to derive the hitting probability for this scenario because it necessitates characterizing the locations of UEs associated with the interfering APs, especially within the context of the rectangular room model. To simplify the antenna gain analysis for interfering APs, we adopt the assumption in \cite{Wu2021TWC} that the depression angle and horizontal beam direction follow uniform distributions.}, depicted in Fig.~\ref{fig:APantenna}. Therefore, the transmit antenna gain equals the main lobe gain with the hitting probability $p_{A} = p_{A,V}p_{A,H}$, where $p_{A,V}=\min\left\{\frac{\phi_{A,V}}{\frac{\pi}{2}-\phi_{AP}},1\right\}$ and $p_{A,H} = \frac{\phi_{A,H}}{2\pi}$ are the probabilities that $U_0$ is located within the AP's vertical beam and horizontal beam, respectively.

\subsubsection{Receive Antenna Gain}
\begin{figure}[t]
    \centering
        \subfigure[The vertical view.]
        {    \label{fig:UEbeamV}        
            \includegraphics[width=0.4\columnwidth]{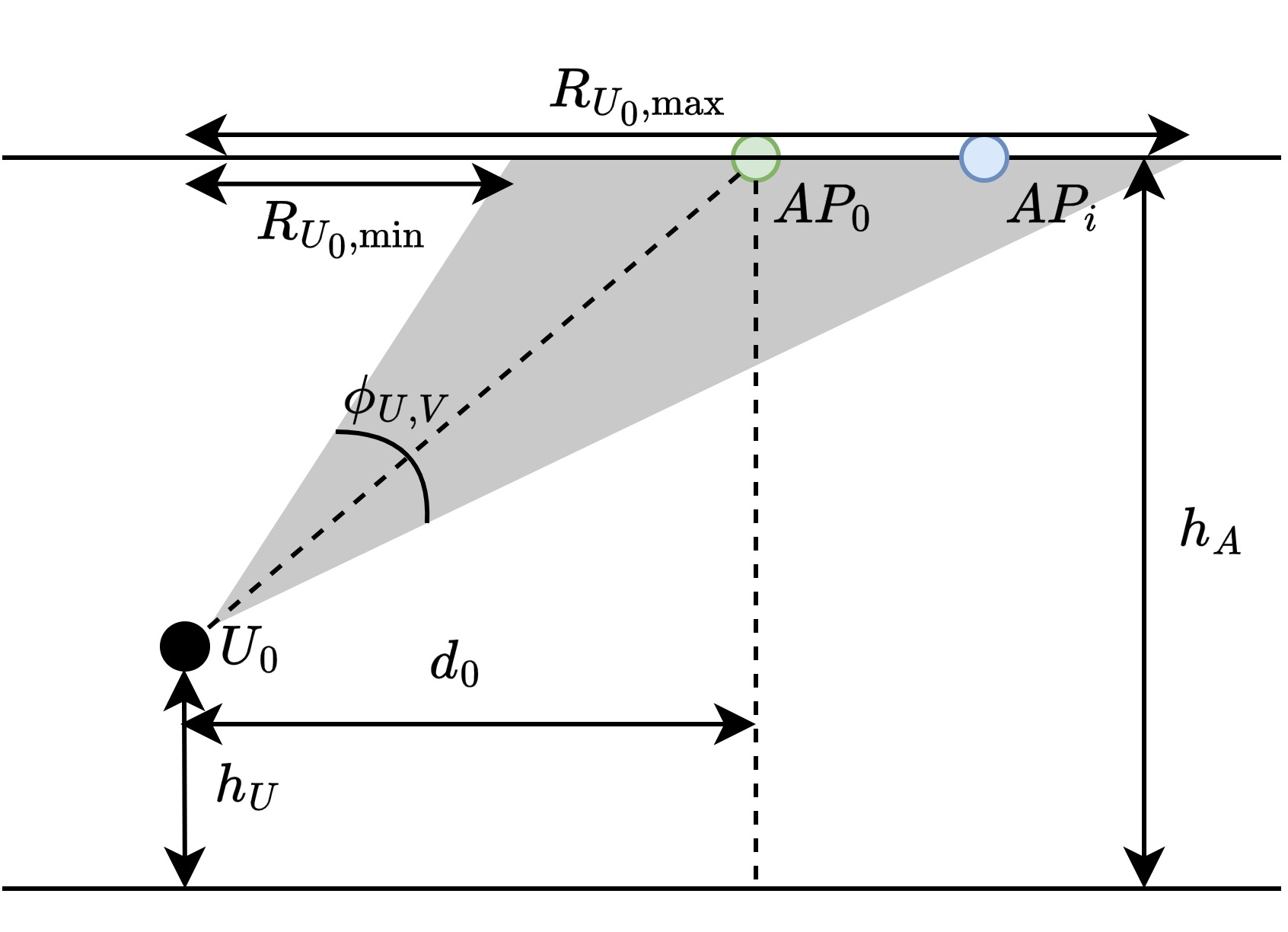}
            }
        \subfigure[The top view.]{\label{fig:UEbeamH}
             \includegraphics[width=0.44\columnwidth]{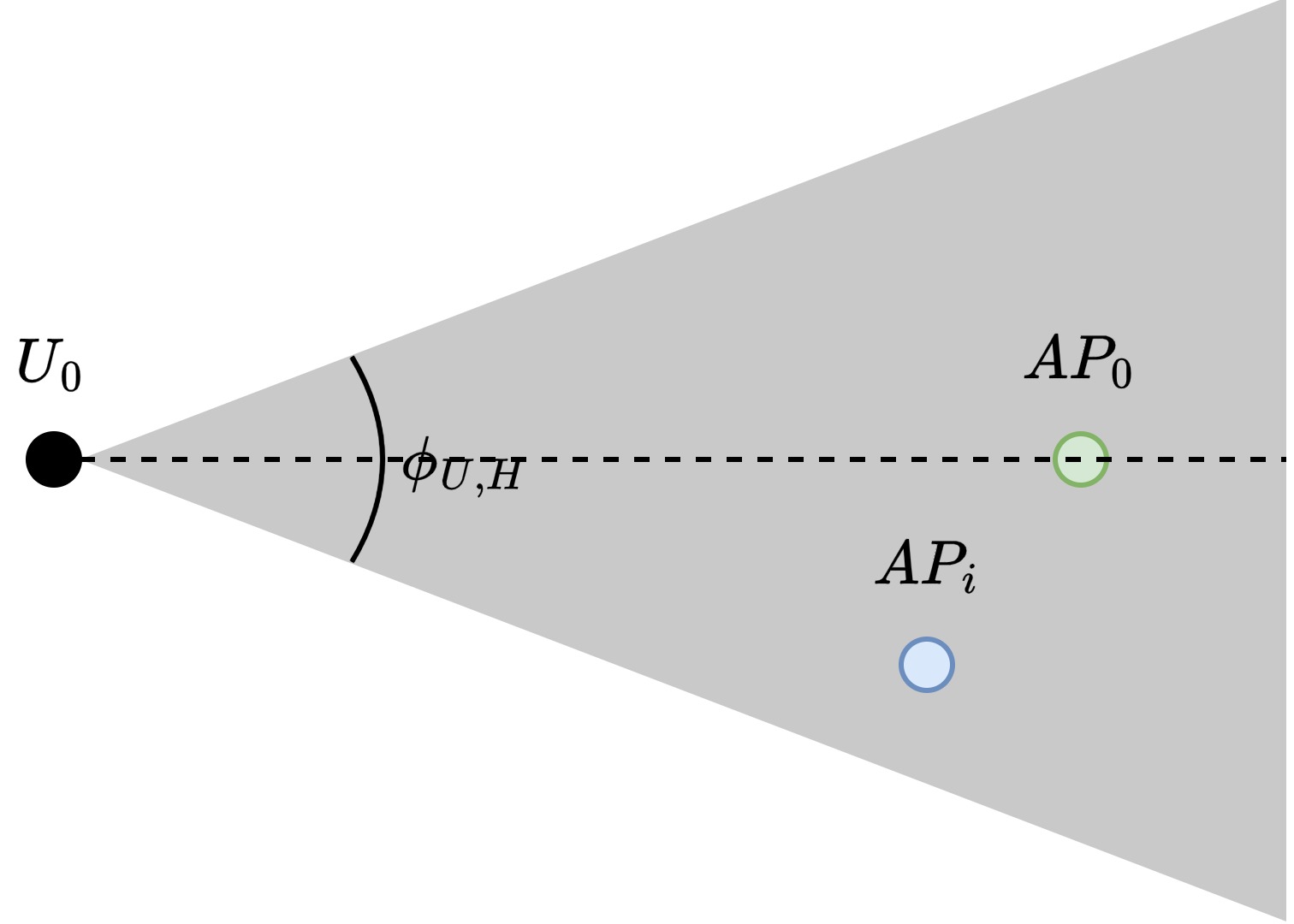}
            }
        \caption{The antenna beam of $U_0$.}
        \vspace{-1.5em}
        \label{fig:UEantenna}
\end{figure}

The receive antenna gain depends on whether the interfering AP is within the antenna beam of $U_0$. We note that the antenna beam of $U_0$ is oriented towards its associated AP, $AP_0$, as depicted in Fig.~\ref{fig:UEantenna}. Similar to the analysis of the transmit antenna gain, the receive antenna gain is equal to the main lobe gain with the hitting probability $p_U = p_{U,V}p_{U,H}$, where $p_{U,V}$ and $p_{U,H}$ are the probabilities that the interfering AP is within the vertical and horizontal beams of $U_0$, respectively.

To determine $p_U$, we first calculate $p_{U,V}$. As illustrated in Fig.~\ref{fig:UEbeamV}, the vertical beam range of $U_0$ is determined by the minimum horizontal distance $R_{U_0,\text{min}}$ and the maximum horizontal distance $R_{U_0,\text{max}}$. According to the nearest LoS AP association strategy, all interfering APs are farther from $U_0$ than $AP_0$. An interfering AP, $AP_i$, is within the vertical beam of $U_0$ if its distance to $U_0$, $d_i$, is less than or equal to $R_{U_0,\text{max}}$. Specifically, $p_{U,V} = 1$ if $d_i\leq R_{U_0,\text{max}}$; otherwise, $p_{U,V} = 0$. The value of $R_{U_0,\text{max}}$ is determined by $d_0$, $h_A$, and $h_U$. If $\arctan\left(\frac{h_A-h_U}{d_0}\right)>\frac{\phi_{U,V}}{2}$, the maximum horizontal distance $R_{U_0,\text{max}}$ is calculated as
\begin{align}
    R_{U_0,\text{max}} = \frac{h_A-h_U}{\tan\left(\arctan\left(\frac{h_A-h_U}{d_0}\right)-\frac{\phi_{U,V}}{2}\right)};
\end{align}
otherwise, $R_{U_0,\text{max}}\rightarrow\infty$, implying that all interfering APs are within the vertical beam of $U_0$. 
We further derive $p_{U,H}$ in the following Lemma.

\begin{Lemma}\label{Lemma:UEinterfer}
    The probability that $AP_i$ is within the horizontal beam of $U_0$ is approximated as
    \begin{align}\label{eq:PUH}
        p_{U,H} \approx \sum\limits_{\vartheta_k\in\Theta(d_0)}\left(\frac{\vartheta_{k}}{\sum\limits_{\vartheta_k\in\Theta(d_0)}\vartheta_{k}}\right)^2p_{\vartheta_k},
    \end{align}
    where
    \begin{align}\label{eq:arcnumber}
        \kappa(d_0) =& \sum\limits_{Z\in\{X,Y\}}\sum\limits_{l\in \{1,2\}}\mathbbm{1}(d_0>R_{Z,l})\notag\\
        &- \sum\limits_{l\in \{1,2\}}\sum\limits_{j \in \{1,2\}}\mathbbm{1}\left(d_0\geq \sqrt{R_{X,l}^2+R_{Y,j}^2}\right)\notag\\
        &+\prod\limits_{Z\in\{X,Y\}}\prod\limits_{l\in \{1,2\}}\mathbbm{1}(d_0<R_{Z,l})
    \end{align}
    denotes the number of arc segments with the center $U_0$ and radius $d_0$ in the room $\mathbf{R}$, $\mathbbm{1}(\cdot)$ is the indicator function, defined as $\1(d_0>R_{Z,l})=1$ if $d_0>R_{Z,l}$; otherwise, $\1(d_0>R_{Z,l})=0$, $\Theta(d_0)=\{\vartheta_1,\cdots,\vartheta_{\kappa(d_0)}\}$ denotes the set of arc segment angles of $\mathrm{ARC}_{d_0,\mathbf{R}}$, given in Table~\ref{tab:1}, and $p_{\vartheta_k}$ is given by
    \begin{align}\label{eq:PRthetak}
       &p_{\vartheta_k} =\notag\\ 
        &\left\{
        \begin{aligned}
            &1, &&\hspace{-30mm}\text{if } 0< \vartheta_{k}\leq\frac{\phi_{UE,H}}{2},\\
           &\frac{\phi_{UE\!,H}}{\vartheta_k}\!-\!\left(\frac{\phi_{UE\!,H}}{2\vartheta_k}\right)^2\!, &&\hspace{-31mm}\text{if }\frac{\phi_{UE,H}}{2}<\vartheta_k\leq 2\pi \!-\!\frac{\phi_{UE,H}}{2},\\
            &\frac{\phi_{UE,H}}{\vartheta_k}+\frac{(\vartheta_k-2\pi)(\vartheta_k-2\pi+\phi_{UE,H})}{\vartheta_k^2}, &&\text{otherwise.}
        \end{aligned}\right.
    \end{align}

    \begin{IEEEproof}
        See Appendix~\ref{Appendix:A}.
    \end{IEEEproof}

\end{Lemma}

\begin{table*}[t]
\centering
\caption{The set of arc segment angles of $\mathrm{ARC}_{d_0,\mathbf{R}}$, $\Theta(d_0)$ .}
\begin{tabular}{|c|c|c|c|c|}
\hline
    $\mathbbm{1}(d_0>R_{X,1})$ & $\mathbbm{1}(d_0>R_{X,2})$ & $\mathbbm{1}(d_0>R_{Y,1})$ & $\mathbbm{1}(d_0>R_{Y,2})$ & $\Theta(d_0)$ \\\hline
     $0$ & $0$ & $0$ & $0$ & \multirow{5}*{$\{\theta(d_0)\}$} \\\cline{1-4}
      $1$ & $0$ & $0$ & $0$ & \multirow{5}*{} \\\cline{1-4}
       $0$ & $1$ & $0$ & $0$ & \multirow{5}*{} \\\cline{1-4}
        $0$ & $0$ & $1$ & $0$ & \multirow{5}*{} \\\cline{1-4}
         $0$ & $0$ & $0$ & $1$ & \multirow{5}*{} \\\hhline{|=|=|=|=|=|}
    $1$ & $1$ & $0$ & $0$ & $\{\theta_{X_1,Y_1}(d_0)+\theta_{X_2,Y_1}(d_0),\theta_{X_1,Y_2}(d_0)+\theta_{X_2,Y_2}(d_0)\}$ \\\hline
    $1$ & $0$ & $1$ & $0$ &  $\{\theta_{X_1,Y_1}(d_0),\theta(d_0)-\theta_{X_1,Y_1}(d_0)\}\setminus\{0\}$\\\hline
    $1$ & $0$ & $0$ & $1$ & $\{\theta_{X_1,Y_2}(d_0),\theta(d_0)-\theta_{X_1,Y_2}(d_0)\}\setminus\{0\}$ \\\hline
    $0$ & $1$ & $1$ & $0$ & $\{\theta_{X_2,Y_1}(d_0),\theta(d_0)-\theta_{X_2,Y_1}(d_0)\}\setminus\{0\}$ \\\hline
    $0$ & $1$ & $0$ & $1$ &  $\{\theta_{X_2,Y_2}(d_0),\theta(d_0)-\theta_{X_2,Y_2}(d_0)\}\setminus\{0\}$\\\hline
    $0$ & $0$ & $1$ & $1$ & $\{\theta_{X_1,Y_1}(d_0)+\theta_{X_1,Y_2}(d_0),\theta_{X_2,Y_1}(d_0)+\theta_{X_2,Y_2}(d_0)\}$ \\\hhline{|=|=|=|=|=|}
    $1$ & $1$ & $1$ & $0$ & $\{\theta_{X_1,Y_1}(d_0),\theta_{X_2,Y_1}(d_0),\theta_{X_1,Y_2}(d_0)+\theta_{X_2,Y_2}(d_0)\}\setminus\{0\}$ \\\hline
    $1$ & $1$ & $0$ & $1$ & $\{\theta_{X_1,Y_1}(d_0)+\theta_{X_2,Y_1}(d_0),\theta_{X_1,Y_2}(d_0),\theta_{X_2,Y_2}(d_0)\}\setminus\{0\}$ \\\hline
    $1$ & $0$ & $1$ & $1$ & $\{\theta_{X_1,Y_1}(d_0),\theta_{X_1,Y_2}(d_0),\theta_{X_2,Y_1}(d_0)+\theta_{X_2,Y_2}(d_0)\}\setminus\{0\}$ \\\hline
    $0$ & $1$ & $1$ & $1$ & $\{\theta_{X_1,Y_1}(d_0)+\theta_{X_1,Y_2}(d_0),\theta_{X_2,Y_1}(d_0),\theta_{X_2,Y_2}(d_0)\}\setminus\{0\}$ \\\hhline{|=|=|=|=|=|}
    $1$ & $1$ & $1$ & $1$ & $\{\theta_{X_1,Y_1}(d_0),\theta_{X_1,Y_2}(d_0),\theta_{X_2,Y_1}(d_0),\theta_{X_2,Y_2}(d_0)\}\setminus\{0\}$ \\\hline
\end{tabular}
\vspace{-1.5em}
\label{tab:1}
\end{table*}



By combining $p_{U,V}$ with $p_{U,H}$, we obtain the hitting probability of an interfering AP, $AP_i$, within the antenna beam of $U_0$. By examining $p_{U,V}$ and $p_{U,H}$ in \eqref{eq:PUH}, we find that this hitting probability monotonically increases with $d_0$. It indicates that a larger distance between a UE and its associated AP results in a higher number of interfering APs falling within the antenna beam of the UE, thereby increasing the overall interference.

Based on our analysis of both transmit and receive antenna gains, the probability distribution of the antenna gain for $AP_i$ is given by
\begin{align}\label{eq:PRGi}
    \mathrm{Pr}(G_i) =\left\{
    \begin{aligned}
        &p_Ap_U, &\text{if }G_i=G_A^m G_U^m,\\
        &p_A(1-p_U), &\text{if }G_i=G_A^m G_U^s,\\
        &(1-p_A)p_U, &\text{if }G_i=G_A^s G_U^m,\\
        &(1-p_A)(1-p_U), &\text{if }G_i=G_A^s G_U^s.
    \end{aligned}
    \right.
\end{align}

\subsection{Coverage Performance Analysis}

Building on the derived $f_{D_0}(d_0)$, interfering AP intensity, and antenna gains for interfering signals, we now analyze the coverage performance of $U_0$. We derive the coverage probability of $U_0$, as presented in the following Theorem.

\begin{Theorem}\label{Theorem:Coverage}
   The coverage probability of $U_0$ is derived as
    \begin{align}\label{eq:PcTheorem}
        P_{c} =& \int_0^{\infty}\frac{m^m}{\Gamma(m)}\sum\limits_{j=0}^{\infty}\frac{K^j r_j}{\Gamma(j+1)}\sum\limits_{l=0}^{j}\frac{(-s)^l}{l!}\notag\\
        &\times\frac{\partial^{(l)} \Lp_{I+N}(s|d_0)}{\partial s^l}\Bigg|_{s=\frac{\beta}{2G_0W(d)\sigma^2}}f(d_0)\mathrm{d}d_0,
    \end{align}
where
\begin{align}\label{eq:LaplaceTheorem}
    &\Lp_{I+N}(s|d_0) = \exp\Bigg(-sN-\int_{d_0}^{\infty}\Lambda_{A}(d)\notag\\
    &\times\Bigg(p_B(d)+(1-p_B(d))\sum\limits_{G_i}\mathrm{Pr}(G_i)\frac{m^m}{\Gamma(m)}\sum\limits_{j=0}^{\infty}\frac{K^j r_j }{j!}\notag\\
    &\times\left(1+2s\sigma^2 \frac{P_t G_{i} c^2W(d)}{(4\pi f)^2}\right )^{-(j+1)}\Bigg)\mathrm{d}d\Bigg),
\end{align}
represents the Laplace transform of the interference plus noise power.

\begin{IEEEproof}
    See Appendix~\ref{Appendix:B}.
\end{IEEEproof}
\end{Theorem}

From Theorem~\ref{Theorem:Coverage}, we find that the antenna gain significantly affects the coverage performance. The coverage performance can be improved by decreasing the antenna beam width and increasing the main lobe gain of both AP's and UE's antennas. Additionally, the coverage probability derived in Theorem~\ref{Theorem:Coverage} can be easily extended to systems incorporating a stochastic wall model, as proposed in \cite{Wu2021TWC}.



\section{Numerical Results}\label{Sec:Num}

\begin{figure}[t]
    \centering
    \includegraphics[width=0.7\columnwidth]{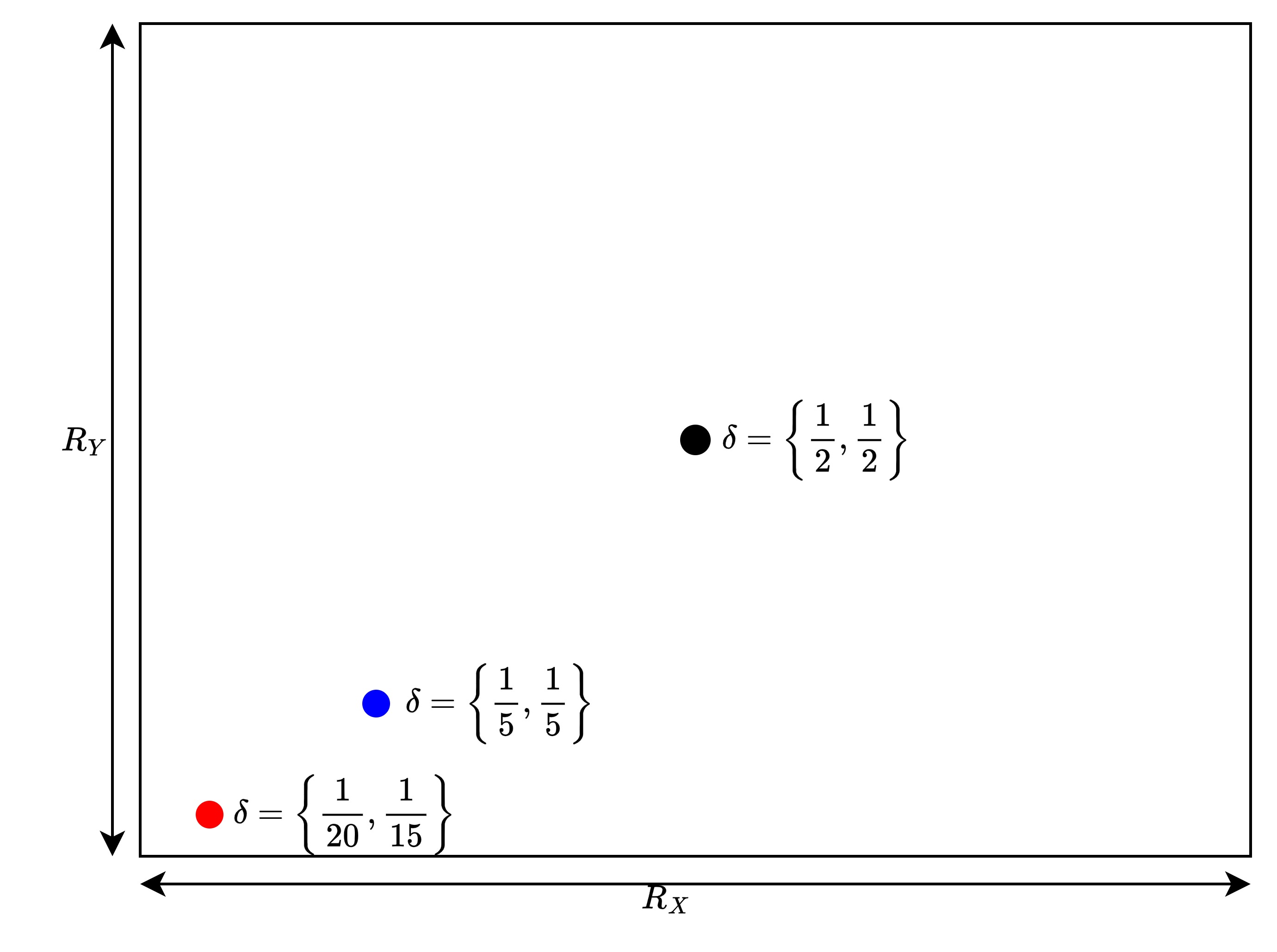}
    \vspace{-0.5em}
    \caption{Illustration of the considered three locations of the typical UE in the room, where the red point is for $\delta=\{\frac{1}{20},\frac{1}{15}\}$, the blue point is for $\delta=\{\frac{1}{5},\frac{1}{5}\}$, and the black point is for $\delta=\{\frac{1}{2},\frac{1}{2}\}$.}
    \vspace{-1em}
    \label{fig:Location}
\end{figure}

In this section, we first present numerical results to validate our analysis in Section~\ref{Sec:Coverage}, including the PDF of the horizontal distance between the typical UE and its associated AP, the hitting probability of UEs, and the coverage probability of the typical UE. We then evaluate the impact of various parameters on the coverage probability, such as the AP density, the location of the typical UE, and the room size. For the location of the typical UE, we consider three scenarios: The corner of the room where $\delta=\{\frac{1}{20},\frac{1}{15}\}$, the near-center of the room where $\delta=\{\frac{1}{5},\frac{1}{5}\}$, and the center of the room where $\delta=\{\frac{1}{2},\frac{1}{2}\}$, as shown in Fig.~\ref{fig:Location}. The simulation results are conducted using MATLAB through averaging over $10^8$ realisations, while the analytical expressions are evaluated using Mathematica. The values of the parameters used in this section are summarized in Table~\ref{tab:System_Para}, unless specified otherwise. These values are similar to values widely used in THz literature \cite{Shafie2021JSAC,Wu2021TWC,Kokkoniemi2020}.

\begin{table*} 
\centering
\caption{Value of System Parameters Used in Section~\ref{Sec:Num}.}
\begin{tabular}{|l|l|l|l|}
\hline
    \textbf{Parameter Type} &\textbf{Parameter} & \textbf{Symbol} & \textbf{Value} \\
    \hline
    
     \multirow{3}*{AP and human blockages}&Density of APs and human blockages  &  $\lambda_A$, $\lambda_B$ & $0.1$ $\mathrm{m}^{-2}$, $0.1$ $\mathrm{m}^{-2}$ \\\cline{2-4}
    \multirow{3}*{}&Radius of human body  &  $R_B$ & $0.25$ $\mathrm{m}$ \\\cline{2-4}
    \multirow{3}*{}&Height of APs, UEs, and human blockages  &  $h_A$, $h_U$, $h_B$ & $3$ m, $1$ m, $1.7$ m\\\hline
    \multirow{2}*{MLP wall blocakge}&Length and width of the room  &  $R_X$, $R_Y$ & $20$ m, $15$ m\\\cline{2-4}
    \multirow{2}*{}&Location of the typical UE  &  $\delta$ & $\{\frac{1}{20},\frac{1}{15}\}$,$\{\frac{1}{5},\frac{1}{5}\}$,$\{\frac{1}{2},\frac{1}{2}\}$\\\hline
    \multirow{4}*{THz transmission} &Operating frequency and  bandwidth &  $f$, $B$ & $300$ GHz, $5$ GHz \\\cline{2-4}
     \multirow{4}*{} &Absorption coefficient &  $K(f)$ & $0.00143$ $\mathrm{m}^{-1}$ \\\cline{2-4}
     \multirow{4}*{}&Transmit power & $P_t$ & $5$ dBm \\\cline{2-4}
    \multirow{4}*{}& AWGN power & $N_0$ & $-77$ dBm \\\hline
    Fading&FTR fading parameters & $K$, $m$, $\sigma$ & $4$, $2$, $1/\sqrt{10}$ \\\hline
    \multirow{2}*{Antenna}&AP's antenna parameters & $G_A^{m}$, $G_{A}^{s}$, $k_A$, $\phi_{A,H}$, $\phi_{A,V}$, $R_A$ & $25$ dBi, $-10$ dBi, $0.1$, $10^{\circ}$,$10^{\circ}$, $20$ m \\\cline{2-4}
    \multirow{2}*{}&UE's antenna parameters & $G_U^{m}$, $G_{u}^{s}$, $k_U$, $\phi_{U,H}$, $\phi_{U,V}$ & $15$ dBi, $-10$ dBi, $0.1$, $33^{\circ}$,$33^{\circ}$ \\\hline
\end{tabular}
\vspace{-1.5em}
\label{tab:System_Para}
\end{table*}

\begin{figure*}[ht!]
    \centering
    \captionsetup[subfigure]{labelfont=footnotesize,textfont=footnotesize}
        \subfigure[For $\delta=\{\frac{1}{2},\frac{1}{2}\}$.]{
            \includegraphics[width=0.31\textwidth]{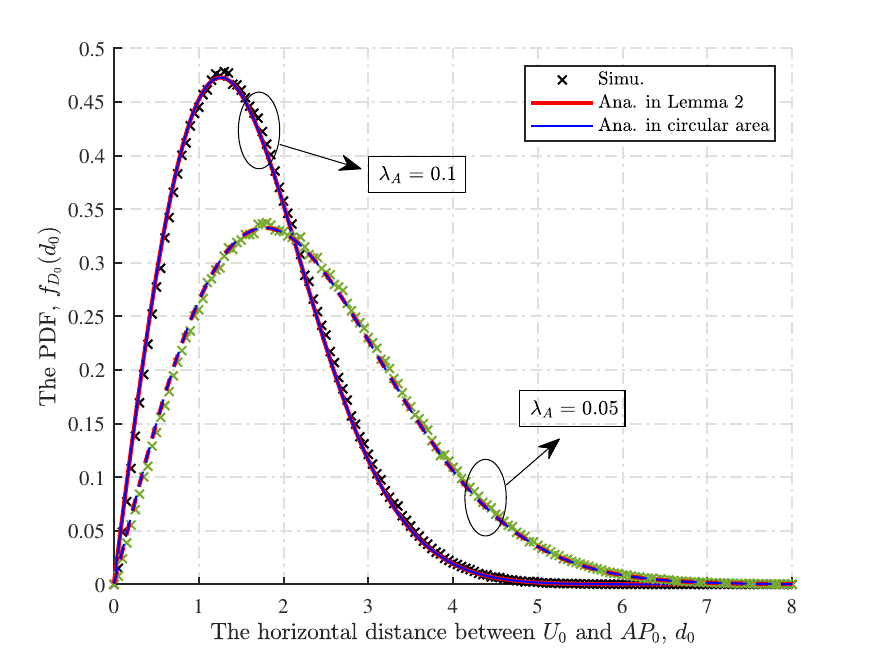}
            }\label{fig:PDF1}
            \hfill
        \subfigure[For $\delta=\{\frac{1}{5},\frac{1}{5}\}$.]{
            \includegraphics[width=0.31\textwidth]{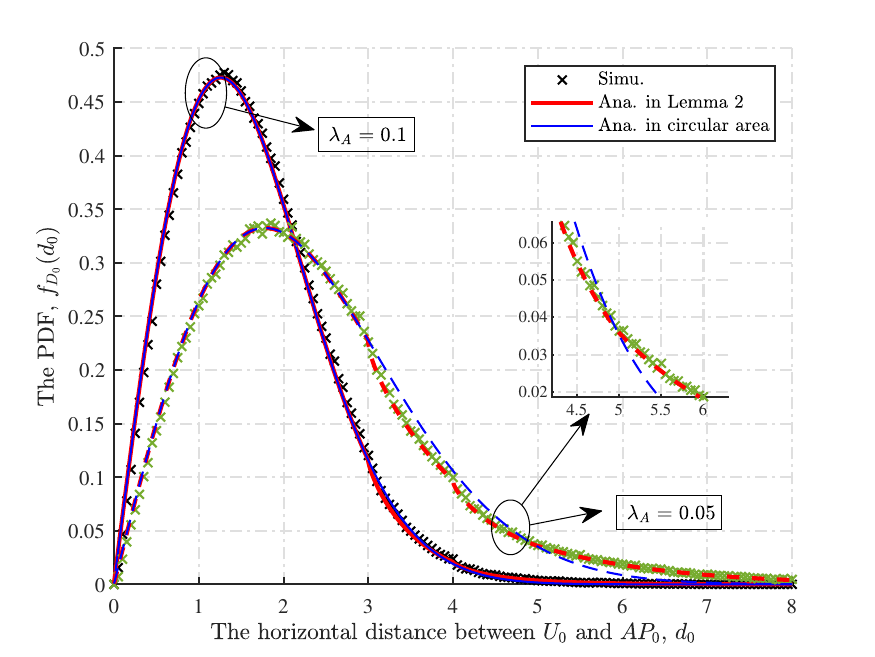}
            }\label{fig:PDF2}
            \hfill
        \subfigure[For $\delta=\{\frac{1}{20},\frac{1}{15}\}$.]{
            \includegraphics[width=0.31\textwidth]{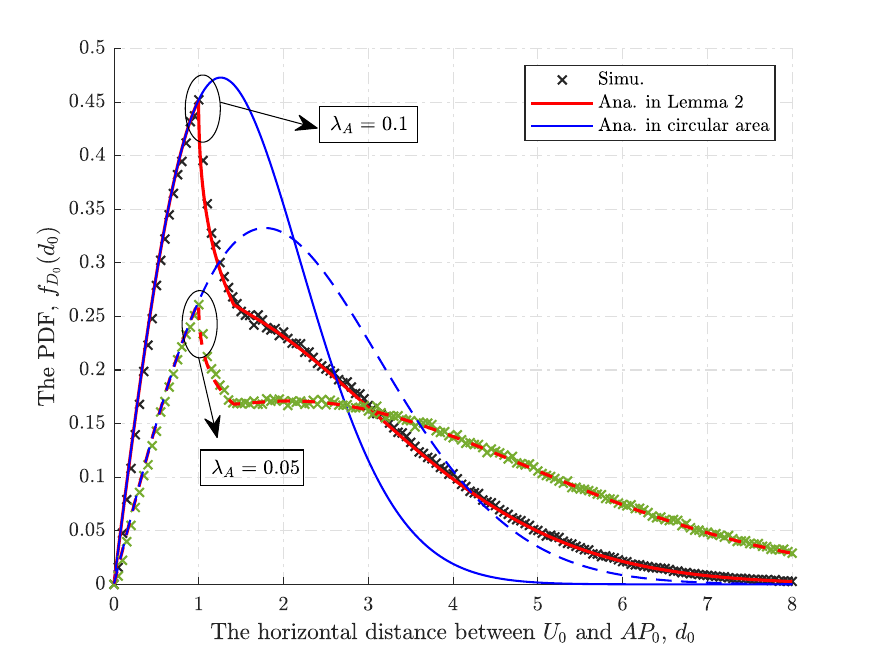}
           } \label{fig:PDF3}
        \caption{The PDF of the horizontal distance between $U_0$ and $AP_0$, $f_{D_0}(d_0)$. The solid line is for $\lambda_A=0.1$ and the dashed line is for $\lambda_A=0.05$.}
        \vspace{-1.5em}
        \label{fig:PDFdis}
\end{figure*}

\subsection{Model Validation}

Fig.~\ref{fig:PDFdis} plots the PDF of the horizontal distance between $U_0$ and  $AP_0$, $f_{D_0}(d_0)$. We first observe that our analytical results in Lemma~\ref{Lemma:intensity} tightly match the simulation results, which demonstrates the correctness of our analysis. Second, we observe that the PDF of the horizontal distance between $U_0$ and $AP_0$ varies for different locations of $U_0$, particularly for a small AP density. This observation is accordance with Remark 1, which further validates our analysis. We then observe that the PDF of the horizontal distance between $U_0$ and $AP_0$ first increases and then decreases. This observation is due to the fact that the increase in the horizontal distance from $U_0$ has a two-fold impact on the PDF of the horizontal distance between $U_0$ and $AP_0$. Specifically, as the horizontal distance from $U_0$ increases, the LoS AP intensity along this distance increases, while the probability of not having a LoS AP along this distance decreases. Furthermore, when comparing our analytical results with those from the circular area model proposed in \cite{Wu2021TWC}, we observe an increasing gap as the UE moves from the center to the corner of the room. Additionally, the PDF of the horizontal distance between $U_0$ and $AP_0$ has a sharp decrease when $U_0$ is located in the corner of the room. This observation suggests that the location of UEs significantly impacts their performance.



\begin{figure}
    \centering
    \includegraphics[width=0.9\columnwidth]{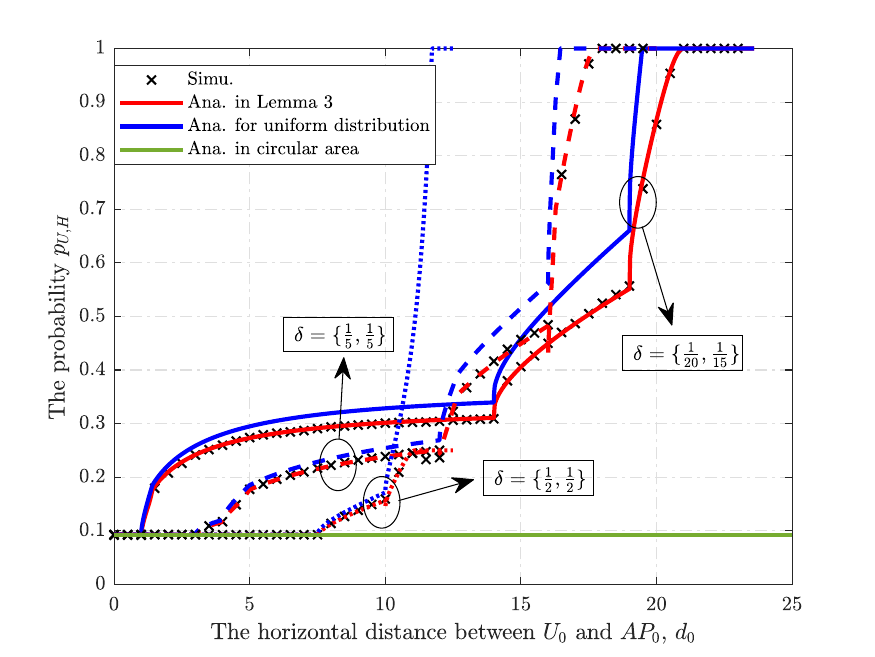}
    \vspace{-0.5em}
    \caption{The probability that an interfering AP is within the horizontal beam of $U_0$, $p_{U,H}$, versus the horizontal distance between $U_0$ and $AP_0$, $d_0$. The solid line is for $\delta=\{\frac{1}{20},\frac{1}{15}\}$, the dashed line is for $\delta=\{\frac{1}{5},\frac{1}{5}\}$, and the dotted line is for $\delta=\{\frac{1}{2},\frac{1}{2}\}$.}
    \vspace{-1.5em}
    \label{fig:hitting}
\end{figure}

Fig.~\ref{fig:hitting} plots the probability that an interfering AP is within the horizontal beam of $U_0$, $p_{U,H}$, versus the horizontal distance between $U_0$ and $AP_0$, $d_0$. We first observe that our analytical results in Lemma~\ref{Lemma:UEinterfer} closely align with the simulation results, validating our analysis. Second, we observe that this probability monotonically increases with $d_0$. This observation is due to the fact that the increase in $d_0$ leads to the decrease in the angle of the arc with the center $U_0$ and radius $d_0$ in the room, limiting the directional possibilities for both the associated AP and interfering APs relative to $U_0$, thereby increasing $p_{U,H}$. Third, we observe that $p_{U,H}$ is significantly underestimated for the circular area, especially when $U_0$ is located in the corner of the room. This is because the horizontal beam of $U_0$ is more likely to be toward the direction of the walls farther from $U_0$, where the number of interfering APs is large, leading to a high probability that an interfeing AP is within the horizontal beam of $U_0$.


\begin{figure}
    \centering
    \includegraphics[width=0.95\columnwidth]{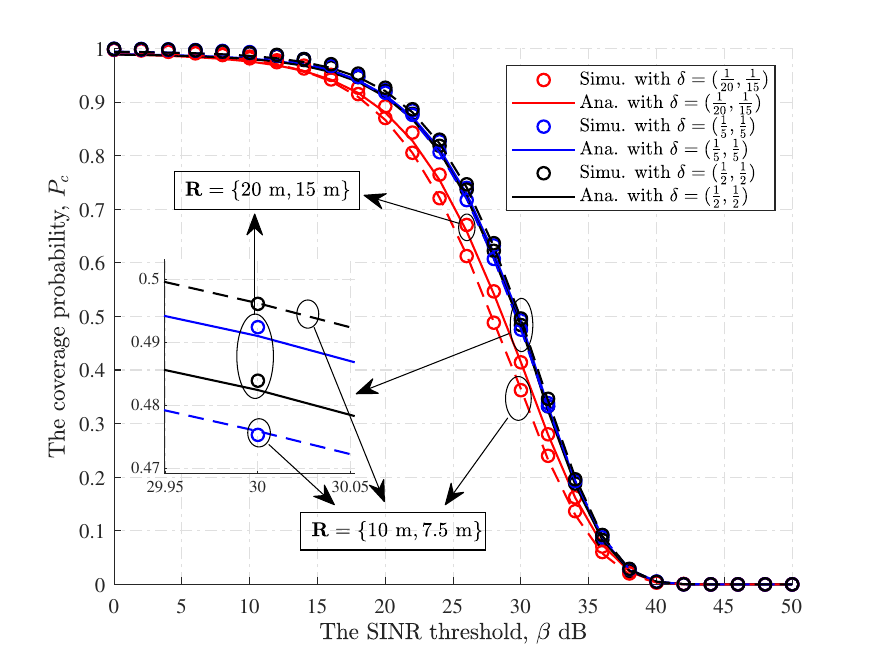}
    \vspace{-0.5em}
    \caption{The coverage probability of $U_0$, $P_c$, versus the the $\mathrm{SINR}$ threshold, $\beta$ dB. The solid line is for $\mathbf{R}=\{20\text{ m, }15\text{ m}\}$ and the dashed line is for $\mathbf{R}=\{10\text{ m, }7.5\text{ m}\}$.}\vspace{-1.8em}
    \label{fig:CoverP}

\end{figure}
\subsection{Impact on Coverage Probability}

Fig.~\ref{fig:CoverP} plots the coverage probability of $U_0$, $P_c$, versus the $\mathrm{SINR}$ threshold, $\beta$. We first observe that our analytical results in Theorem~\ref{Theorem:Coverage} tightly match the simulation results, validating the accuracy of our analysis. We then observe that $P_c$ is lower when $U_0$ is located in the corner of the room compared to at the center. This observation is due to the fact that both the average distance between $U_0$ and $AP_0$ and the hitting probability of $U_0$ are higher when $U_0$ is located in the corner of the room compared to at the center. When $U_0$ associates with a more distant AP, the received signal power decreases, and more interfering APs are within the beam of $U_0$, which degrades the coverage performance.


\begin{figure}
    \centering
    \includegraphics[width=0.9\columnwidth]{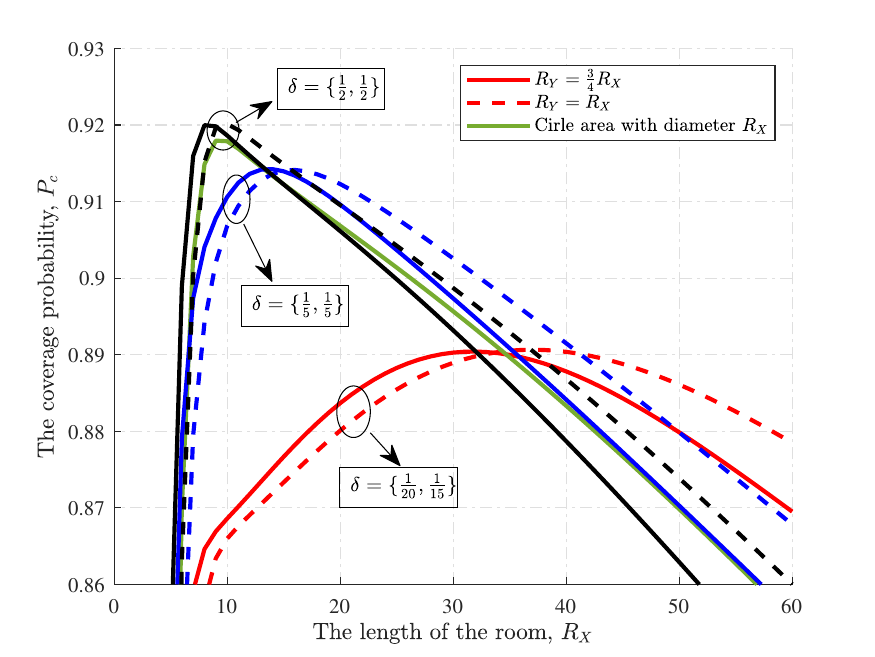}
    \vspace{-0.5em}
    \caption{The coverage probability of $U_0$, $P_c$, versus the length of the room, $R_X$, with $\beta=20$ dB. The red line is for $\delta=\{\frac{1}{20},\frac{1}{15}\}$, the blue line is for $\delta=\{\frac{1}{5},\frac{1}{5}\}$, and the black line is for $\delta=\{\frac{1}{2},\frac{1}{2}\}$.}
    \vspace{-1.5em}
    \label{fig:CPvsRoom}
\end{figure}

Fig.~\ref{fig:CPvsRoom} plots the coverage probability of $U_0$, $P_c$, versus the length of the room, $R_X$, with $\beta=20$ dB. We first observe that the coverage probability dramatically increases and then decreases as the room size increases. This observation is due to the fact that the increase in the room size results in a two-fold effect on the coverage probability. In particular, when the room size increases from a small value, the increasing LoS AP association probability of UEs dominantly and positively affects the coverage probability. When the room size exceeds a certain threshold, the number of interfering APs increases significantly, thereby degrading the coverage probability. Moreover, this effect is more pronounced when $U_0$ is located at the center of the room, compared to in the corner. 
Furthermore, we observe that the coverage probability derived for the circular area model is close to that for a UE located at the center of the room but significantly differs for a UE located in the corner. This observation indicates that while the analysis for a circular area model can be adapted to simplify the analysis for UEs at the center, it is not suitable for UEs located in the corner.

\begin{figure}
    \centering
    \includegraphics[width=0.95\columnwidth]{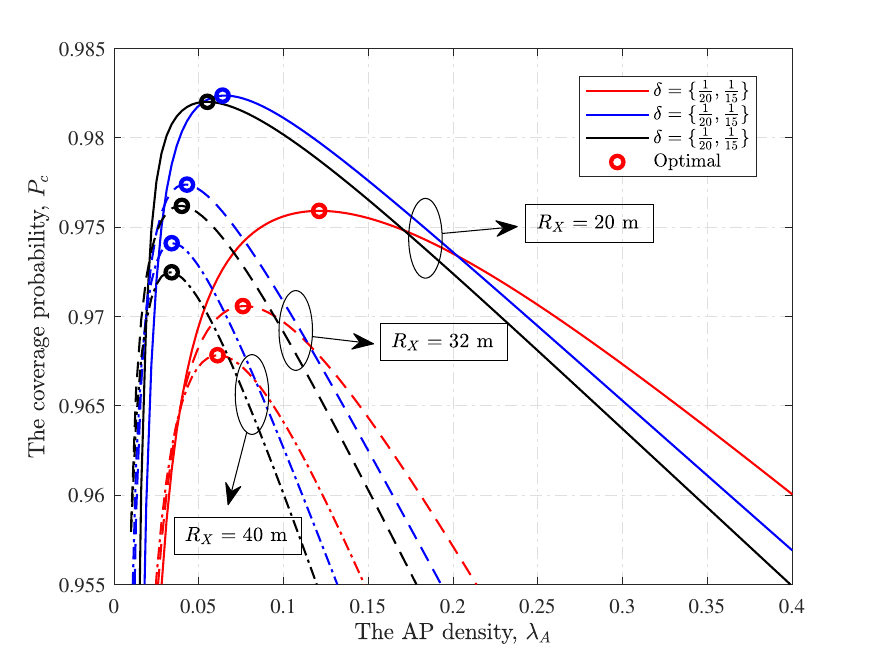}
    \vspace{-0.5em}
    \caption{The coverage probability of $U_0$, $P_c$, versus the density of AP, $\lambda_A$, with $\beta=10$ dB and $R_Y=\frac{3}{4}R_X$. The solid line is for $R_X=20$ m, the dashed line is for $R_X=32$ m, and the dash-dotted line is for $R_X=40$ m.}
    \vspace{-1em}
    \label{fig:CoPvsLa}
\end{figure}

Fig.~\ref{fig:CoPvsLa} plots the coverage probability of $U_0$, $P_c$, versus the AP density AP, $\lambda_A$, with $\beta=10$ dB and $R_Y=\frac{3}{4}R_X$. We first observe that the coverage probability first increases and then decreases as $\lambda_A$ increases. This observation is because the AP density has a two-fold effect on the coverage probability. When $\lambda_A$ is small, its increase reduces the distance between $U_0$ and its associated AP, thereby increasing the received signal power and the coverage probability. However, once $\lambda_A$ exceeds a certain threshold, its increase leads to a significant increase in the number of interfering APs, which results in a higher interference and, consequently, a decrease in the coverage probability. This observation also indicates that there exists an optimal AP density to maximize the coverage probability.
Moreover, we find that the location of UEs has a significant impact on this optimal AP density, with nearly double the density required when the UE is in the corner of the room compared to the center. 
This insight is crucial for designing future THz systems in indoor environments to meet coverage requirements for all UEs in the room. Furthermore, we observe that the impact of increasing $\lambda_A$ on the coverage probability is more pronounced when $U_0$ is at the center of the room compared to in the corner. This is because the increase in $\lambda_A$ results in more interfering APs for a UE located at the center, resulting in significantly higher interference.


\section{Conclusion}\label{Sec:Conclusion}

In this work, we developed a tractable analytical framework to assess the coverage performance of a typical UE in an indoor THz communication system. We began by modeling a realistic 3D THz system, incorporating the unique molecular absorption loss and small-scale fading at THz frequencies, and 3D directional antennas at both UEs and APs. Differing from existing studies that used a Boolean scheme of straight-line wall blockage model, we modeled the wall blockages by MLPs and the human blockages by a random circle process. With these blockage models, we analyzed the impact of a UE's location on the distance to its associated AP and the intensity of interfering APs. We then derived hitting probabilities that form the foundation of the coverage analysis. Thereafter, we derived a new expression for the coverage probability of the typical UE by incorporating a precise THz fading model, i.e., the FTR channel model. Using numerical results, we first validated our analysis and examined how the UE's location affects its AP association and received interference. We then discovered that the optimal AP density for maximizing the coverage probability is determined by the UE's location and the room size. This insight lays a crucial foundation for designing future THz communication systems in indoor environments to meet coverage requirements.

\begin{appendices}
\section{Proof of Lemma \ref{Lemma:arc}}\label{Appendix:Lemma1}
 According to the location of $U_0$, the horizontal plane of the room can be divided into four separate rectangular spaces with lengths $R_{X,i}$ and widths $R_{Y,j}$, where $i\in\{1,2\}$ and $j\in\{1,2\}$. We note that the angle of $\mathrm{ARC}_{d,\mathbf{R}}$ can be calculated by summarizing it from each rectangular space as
    \begin{align}\label{eq:thetadsum1}
        \theta(d) = \sum\limits_{i\in\{1,2\}}\sum\limits_{j\in\{1,2\}}\theta_{X_i,Y_j}(d),
    \end{align}
    where $\theta_{X_i,Y_j}(d)$ denotes the angle of $\mathrm{ARC}_{d,\mathbf{R}}$ in the rectangular space with the length $R_{X,i}$ and width $R_{Y,j}$. Without loss of generality, we analyze the angle of $\mathrm{ARC}_{d,\mathbf{R}}$ in the rectangular space with $R_{X,1}$ and $R_{Y,1}$, as shown in Fig.~\ref{fig:RoomOnePart}. If $d< \sqrt{R_{X,1}^2+R_{Y,1}^2}$, $\theta_{X_1,Y_1}(d) $ is calculated as
    \begin{align}
        \theta_{X_1,Y_1}(d) = \frac{\pi}{2}-\psi_{R_{X,1},d}-\psi_{R_{Y,1},d};
    \end{align}
    otherwise, $\theta_{X_1,Y_1}(d) = 0$. We note that $\psi_{R_{Z,1},d} = \arccos\left(\frac{R_{Z,1}}{d}\right)$, where $Z\in\{X,Y\}$, if $R_{Z,1}>d$; otherwise, $\psi_{R_{Z,1},d}=0$, which leads to \eqref{eq:psiab}. Combining $\theta_{X_1,Y_1}(d)$ for $d< \sqrt{R_{X,1}^2+R_{Y,1}^2}$ and $d\geq\sqrt{R_{X,1}^2+R_{Y,1}^2}$, we obtain
    \begin{align}\label{eq:thetaRx1RY1}
        \theta_{X_1,Y_1}(d) = \left(\frac{\pi}{2} - \psi_{R_{X,1},{d}} - \psi_{R_{Y,1},{d}}\right)^+.
    \end{align}
    By substituting \eqref{eq:thetaRx1RY1} into \eqref{eq:thetadsum1}, we obtain the angle of $\mathrm{ARC}_{d,\mathbf{R}}$ 
    as \eqref{eq:arcradius}. Moreover, based on the formula for the arc length, we obtain \eqref{eq:arclength}.

\section{Proof of Lemma \ref{Lemma:intensity}}\label{Appendix:Lemma2}
Based on the human blockage model, the LoS AP intensity located in the room $\mathbf{R}$ at a horizontal distance $d$ from $U_0$ is given by
    \begin{align}
        \Lambda_{A,\mathrm{LoS}}(d) = e^{-\alpha d }\Lambda_{A}(d).
    \end{align}
According to the property of the PPP, the probability that the number of LoS APs within the distance $d$ from $U_0$, denoted by $N_{A,\mathrm{LoS}}(d)$, equals $n$ is given by
\begin{align}
    \mathrm{Pr}\left(N_{A,\mathrm{LoS}}(d)=n\right) = \frac{(\rho(d))^n}{n!}\exp(-\rho(d)),
\end{align}
where $\rho(d)=\int_{0}^{d}\Lambda_{A,\mathrm{LoS}}(x) \mathrm{d}x$. Thus, the CDF of $d_0$ is given by
\begin{align}\label{eq:FD0}
    F_{D_0}(d_0) = 1\!-\!\mathrm{Pr}\left(N_{A,\mathrm{LoS}}(d_0)\!=\!0\right)\! =\! 1\!-\!\exp(-\rho(d_0)).
\end{align}
We then obtain $f_{D_0}(d_0)$ by taking the derivative of $F_{D_0}(d_0)$, which is given by $f_{D_0}(d_0) = \frac{\mathrm{d}F_{D_0}(d_0)}{\mathrm{d}d_0}$, resulting in \eqref{eq:pdfLoSAP}.

\section{Proof of Lemma \ref{Lemma:UEinterfer}}\label{Appendix:A}

\begin{figure}[t]
    \centering
    \includegraphics[width=0.6\columnwidth]{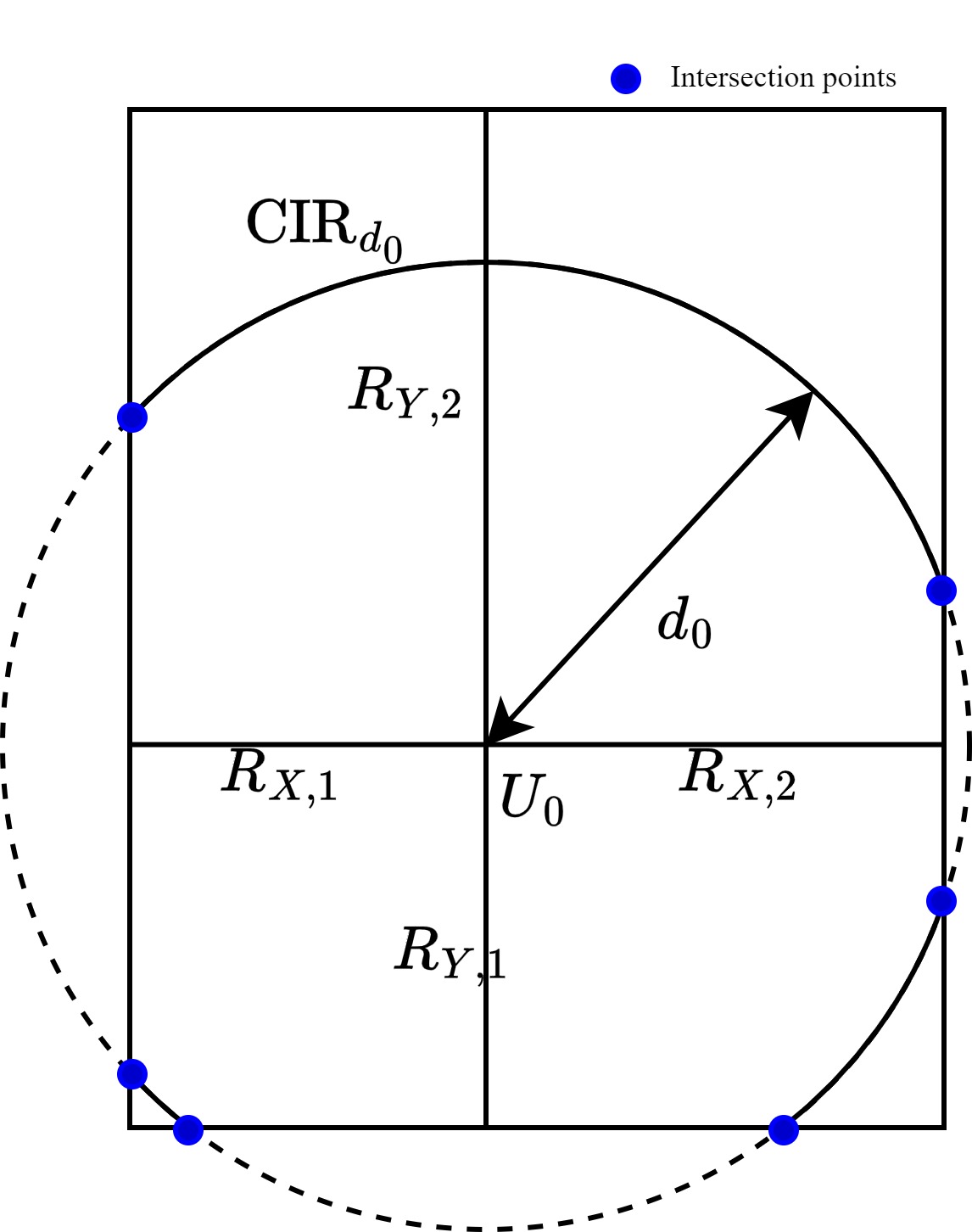}
   \vspace{-0.5em}
    \caption{An example to show the relationship between the number of arc segments of $\mathrm{CIR}_{d_0}$ and the number of intersection points where the circle $\mathrm{CIR}_{d_0}$ intersects the walls of room $\mathbf{R}$, $\xi(d_0)$.}
    \vspace{-1em}
    \label{fig:ArcCirep}
\end{figure}

To calculate $p_{U,H}$, we first analyze the impact of the locations of $AP_0$ and interfering APs. The walls of the room $\mathbf{R}$ divide $\mathrm{ARC}_{d_0,\mathbf{R}}$ into multiple arc segments, as shown in Fig.~\ref{fig:RoomR}. We denote $\mathrm{CIR}_{d_0}$ as the entire circle with the center $U_0$ and radius $d_0$. Since each intersection point where $\mathrm{CIR}_{d_0}$ intersects the walls of room $\mathbf{R}$ divides this circle into arc segments, the total number of arc segments of $\mathrm{CIR}_{d_0}$ is equal to the number of intersection points, as shown in Fig.~\ref{fig:ArcCirep}. We denote $\xi_{X,1}(d_0)$, $\xi_{X,2}(d_0)$, $\xi_{Y,1}(d_0)$, and $\xi_{Y,2}(d_0)$ as the number of intersection points where $\mathrm{CIR}_{d_0}$ intersects the left, right, front, and rear walls, respectively. These values can be calculated by
\begin{align}
    \xi_{X,l}(d_0) \!= \!2\!\times\!\mathbbm{1}(d_0\!>\!R_{X,l})\!- \!\sum\limits_{j\in \{1,2\}}\mathbbm{1}\left(d_0\!\geq\! \sqrt{R_{X,l}^2\!+\!R_{Y,j}^2}\right)
\end{align}
and 
\begin{align}
    \xi_{Y,l}(d_0)\! =\! 2\!\times\!\mathbbm{1}(d_0\!>\!R_{Y,l})\!-\! \sum\limits_{j\in \{1,2\}}\mathbbm{1}\left(d_0\!\geq\! \sqrt{R_{Y,l}^2\!+\!R_{X,j}^2}\right),
\end{align}
for $l\in\{1,2\}$.
Thus, the total number of intersection points where $\mathrm{CIR}_{d_0}$ intersects all the walls of the room $\mathbf{R}$, denoted by $\xi(d_0)$, is calculated as
\begin{align}
    \xi(d_0) &= \sum\limits_{Z\in\{X,Y\}}\sum\limits_{l\in \{1,2\}}\xi_{Z,l}(d_0)\notag\\
    &= 2\sum\limits_{Z\in\{X,Y\}}\sum\limits_{l\in \{1,2\}}\mathbbm{1}(d_0>R_{Z,l}) \notag\\
    &- 2\sum\limits_{l\in \{1,2\}}\sum\limits_{j \in \{1,2\}}\mathbbm{1}\left(d_0\geq \sqrt{R_{X,l}^2+R_{Y,j}^2}\right).
\end{align}

We note that the number of arc segments, $\kappa(d_0)$, of $\mathrm{ARC}_{d_0,\mathbf{R}}$ is equal to half of $\xi(d_0)$, since the number of arc segments in the room $\mathbf{R}$ equals to the number of arc segments outside the room $\mathbf{R}$ if $\xi(d_0)\neq0$. In addition, when $\xi(d_0)=0$, $\kappa(d_0)=1$ if the entire circle is in the room $\mathbf{R}$, i.e., $d_0<R_{Z,j}$ for all $Z\in\{X,Y\}$ and $j\in\{1,2\}$; otherwise, $\kappa(d_0)=0$. Thus, we obtain the number of arc segments within $\mathbf{R}$ as given in \eqref{eq:arcnumber}.

The angle of each arc segment is a combination of angles of $\mathrm{ARC}_{d_0,\mathbf{R}}$ in the rectangular space with the length
$R_{X,l}$ and width $R_{Y,j}$, i.e., $\theta_{X_l,Y_j}(d_0)$, where $l\in\{1,2\}$ and $j\in\{1,2\}$. For example, as shown in Fig.~\ref{fig:ArcCirep}, we have $\Theta(d_0)=\{\theta_{X_1,Y_1}(d_0),\theta_{X_2,Y_1}(d_0),\theta_{X_1,Y_2}(d_0)+\theta_{X_2,Y_2}(d_0)\}$, where $\theta_{X_1,Y_2}(d_0)$ and $\theta_{X_2,Y_2}(d_0)$ are combined to form the third element in $\Theta(d_0)$. This combination depends on whether the arc segments in adjacent rectangular spaces with lengths $R_{X,l}$ and widths $R_{Y,j}$ are connected, determined by the value of $\mathbbm{1}(d_0>R_{Z,l})$. Based on this, $\Theta(d_0)$ is presented in Table~\ref{tab:1}.

\begin{figure}[t]
    \centering
    \includegraphics[width=0.6\columnwidth]{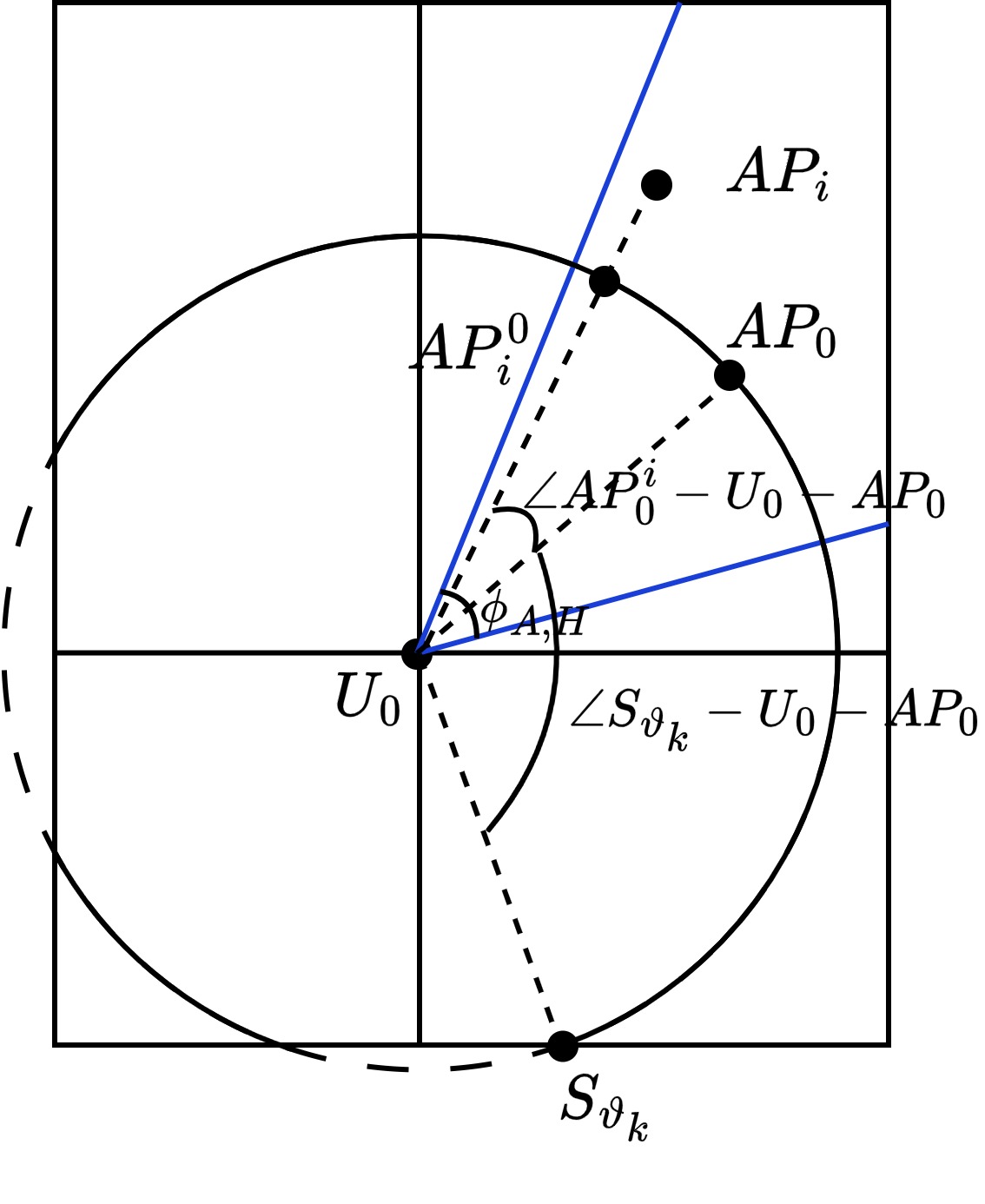}
     \vspace{-1.5em}
    \caption{The interfering AP, $AP_i$, is within the horizontal beam of $U_0$.}
\vspace{-1.5em}
    \label{fig:Antennai}
\end{figure}

The event that $AP_i$ is located within the horizontal beam of $U_0$ is equivalent to $AP_{i}^0$ being within the horizontal beam of $U_0$, where $AP_{i}^0$ is the intersection point of the line connecting $AP_i$ and $U_0$ with the arc $\mathrm{APC}_{d_0,\mathbf{R}}$, as shown in Fig.~\ref{fig:Antennai}.
Here, we assume that $AP_{i}^0$ is located in the horizontal beam of $U_0$ only when $AP_{i}^0$ and $AP_0$ are on the same arc of $\mathrm{ARC}_{d_0,\mathbf{R}}$. According to the property of the PPP, the probability that $AP_{i}^0$ and $AP_0$ are on the same arc with angle $\vartheta_k$ is given by
\begin{align}\label{eq:PrthatakTwo}
    \mathrm{Pr}(\vartheta_k) = \left(\frac{\vartheta_{k}}{\sum\limits_{\vartheta_k\in\Theta(d_0)}\vartheta_{k}}\right)^2.
\end{align}
We denote $\angle AP_{i}^0U_0AP_0$ as the angle formed by lines $U_0-AP_{i}^0$ and $U_0-AP_0$. We then denote $S_{\vartheta_k}$ as one of the intersection points of the arc segment with angle $\vartheta_k$ with the walls of the room $\mathbf{R}$ and $\angle S_{\vartheta_k}^0U_0AP_0$ as the angle formed by lines $U_0-S_{\vartheta_k}$ and $U_0-AP_0$. We note that $AP_{i}^0$ is located in the horizontal beam of $U_0$ if $\angle AP_{i}^0U_0AP_0$ is less than $\phi_{U,H}/2$. The probability that $AP_{i}^0$ is located in the horizontal beam of $U_0$ given that $AP_{i}^0$ and $AP_0$ are on the same arc with angle $\theta_k$, denoted by $p_{\vartheta_k}$, is calculated by
\begin{align}
    p_{\vartheta_k} = &\int_{0}^{\vartheta_k}\mathrm{Pr}\left(\angle AP_{i}^0U_0AP_0<\frac{\phi_{U,H}}{2}\right)\notag\\
    &\times f(\angle S_{\vartheta_k}U_0AP_0)\mathrm{d}\angle S_{\vartheta_k}U_0AP_0.
\end{align}
Due to the uniform distribution of $\angle S_{\vartheta_k}U_0AP_0$ over $[0,\vartheta_k]$, we obtain $p_{\vartheta_k}$ as shown in \eqref{eq:PRthetak}. By combining \eqref{eq:PRthetak} and \eqref{eq:PrthatakTwo}, we obtain \eqref{eq:PUH}.

\section{Proof of Theorem \ref{Theorem:Coverage}}\label{Appendix:B}


Based on \eqref{eq:Pc}, the coverage probability of $U_0$ can be derived by using $\mathrm{Pr}(\mathrm{SINR}>\beta|d_0)$ and $f_{D_0}(d_0)$, where $f_{D_0}(d_0)$ is given in \eqref{eq:pdfLoSAP}. The conditional coverage probability $\mathrm{Pr}(\mathrm{SINR}>\beta|d_0)$, given the horizontal distance $d_0$ between $U_0$ and $AP_0$, is expressed as
\begin{align}\label{eq:PrH0geqT1}
    &\mathrm{Pr}(\mathrm{SINR}>\beta|d_0) = \mathrm{Pr} (g_0W(d_0)H_0>\beta|d_0)\notag\\
    &= \E_I\left[\mathrm{Pr}\left(H_0>\frac{\beta}{g_0 W(d_0)}(I +N)\Bigg|d_0,I\right)\right].
\end{align}
Using the CDF of $H_0$ in \eqref{eq:cdfFTR}, we obtain
\begin{align}\label{eq:PrH0geqT}
    &\mathrm{Pr}\left(H_0>\frac{\beta}{g_0 W(d_0)}(I +N)\Bigg|d_0,I\right)\notag\\
    &=1 - F_h\left(\frac{\beta}{g_0 W(d_0)}(I +N)\Bigg|d_0,I\right)\notag\\
    &=\frac{m^m}{\Gamma(m)}\!\sum\limits_{j\!=\!0}^{\infty}\!\frac{K^j r_j}{\Gamma(j\!+\!1)}\sum\limits_{l\!=\!0}^{j}\frac{\left(\frac{\beta(I+N)}{2g_0W(d_0)\sigma^2}\right)^l}{l!}e^{-\frac{\beta(I+N)}{2g_0W(d_0)\sigma^2}}\Bigg|d_0, I.
\end{align}
By substituting \eqref{eq:PrH0geqT} into \eqref{eq:PrH0geqT1}, we obtain
\begin{align}\label{eq:PrLaplace}
    &\mathrm{Pr}(\mathrm{SINR}>\beta|d_0) \notag\\
    &=\frac{m^m}{\Gamma(m)}\sum\limits_{j=0}^{\infty}\frac{K^j r_j}{\Gamma(j+1)}\notag\\
    &\times\sum\limits_{l=0}^{j}\E_I\left[\frac{\left(\frac{\beta(I+N)}{2g_0W(d_0)\sigma^2}\right)^l}{l!}e^{-\frac{\beta(I+N)}{2g_0W(d_0)\sigma^2}}\Bigg|d,I\right]\notag\\
    &=\frac{m^m}{\Gamma(m)}\!\sum\limits_{j=0}^{\infty}\frac{K^j r_j}{\Gamma(j\!+\!1)}\!\sum\limits_{l=0}^{j}\frac{(-s)^l}{l!}\!\frac{\partial^{(l)} \Lp_{I\!+\!N}(s|d_0)}{\partial s^l}\Bigg|_{s\!=\!\frac{\beta}{2g_0W(d_0)\sigma^2}}.
\end{align}
We then calculate $\Lp_{I+N}(s|d_0)$ in \eqref{eq:PrLaplace} as
\begin{align}\label{eq:LaplaceIN}
    \Lp_{I+N}(s|d_0)&= \E[e^{-s(I+N)}|d_0]\notag\\
    &=e^{-sN_0}\E\left[\exp\left(-s\sum\limits_{AP_i\in\Psi_{AP}}P_i\right)\Bigg|d_0\right].
\end{align}
According to the property of the PPP \cite{Tang2020}, the second term in \eqref{eq:LaplaceIN} is derived as
\begin{align}\label{eq:Espgivend0}
    &\E\left[\exp\left(-s\sum\limits_{AP_i\in\Psi_{AP}}P_i\right)\Bigg|d_0\right] \notag\\
    &= \E\left[\prod\limits_{AP_i\in\Psi_{AP}}\E\left[\exp\left(-s P_i\right)\bigg|d_0\right]\right]\notag\\
    &=\exp\left(-\int_{d_0}^{\infty}(1-\E\left[\exp\left(-s P_i\right)\right])\lambda_{A,LoS}(d)\mathrm{d}d\right),
\end{align}
where $\E\left[\exp\left(-s P_i\right)\right]$ is computed as
\begin{align}\label{eq:Espidid}
    &\E\left[\exp\left(-s P_i\right)\right]\notag\\
    &=p_B(d_i)+(1-p_B(d_i))\E_{g_i,H_i}[\exp(-sg_i W(d_i)H_{i})]\notag\\
    &=p_B(d_i)+(1-p_B(d_i))\sum\limits_{G_i}\mathrm{Pr}(G_i)\notag\\
    &\times \E_{H_i}\left[\exp\left(-s\frac{P_t G_{i} c^2W(d_i)}{(4\pi f)^2}H_{i}\right)\right]\notag\\
    &=p_B(d_i)+(1-p_B(d_i))\sum\limits_{G_i}\mathrm{Pr}(G_i)\frac{m^m}{\Gamma(m)}\notag\\
    &\times\sum\limits_{j=0}^{\infty}\frac{K^j r_j }{j!}\left(1+2s\sigma^2 \frac{P_t G_{i} c^2W(d_i)}{(4\pi f)^2}\right )^{-(j+1)}.
\end{align}
By substituting \eqref{eq:Espgivend0} into \eqref{eq:LaplaceIN}, we obtain $\Lp_{I+N}(s|d_0)$ in \eqref{eq:LaplaceTheorem}. Finally, by substituting \eqref{eq:LaplaceTheorem} into \eqref{eq:PrLaplace} and combining it with \eqref{eq:pdfLoSAP}, we obtain the coverage probability of $U_0$ as \eqref{eq:PcTheorem}.
\end{appendices}

\bibliographystyle{IEEEtran} 
\bibliography{bibli}

\end{document}